\documentclass[aps,prl,reprint,amssymb,showpacs,twocolumns,superscriptaddress,notitlepage]{revtex4-2} 
\usepackage{bm}
\usepackage{graphicx}
\usepackage{dcolumn}
\usepackage{braket}
\usepackage{bbold}
\usepackage{natbib}
\usepackage{amsmath}
\usepackage{color}
\usepackage{dsfont}
\usepackage{comment}
\usepackage{ragged2e}
\usepackage{tikz}
\usepackage{adjustbox}

\definecolor{mayablue}{rgb}{0.45, 0.76, 0.98}
\definecolor{deepskyblue}{rgb}{0.0, 0.75, 1.0}
\definecolor{dodgerblue}{rgb}{0.12, 0.56, 1.0}
\definecolor{ultramarineblue}{rgb}{0.25, 0.4, 0.96}
\definecolor{portlandorange}{rgb}{1.0, 0.35, 0.21}
\definecolor{purple(x11)}{rgb}{0.63, 0.36, 0.94}
\definecolor{gold}{rgb}{0.99, 0.76, 0.0}

\definecolor{forestgreen}{RGB}{34,139,34}

\usepackage[colorlinks]{hyperref}
\hypersetup{   
	linkcolor=black, 
	urlcolor=blue,
    citecolor=blue,
    filecolor=blue,
	pdfstartview=
}
\usepackage[normalem]{ulem}

\usepackage[mathscr]{euscript}

\usepackage{orcidlink}

\usepackage{enumitem}

\usepackage{lipsum}
\newbox\one
\newbox\two
\long\def\loremlines#1{%
    \setbox\one=\vbox {%
      \lipsum%
     }
   \setbox\two=\vsplit\one to #1\baselineskip
   \unvbox\two}

\usepackage[titles]{tocloft} 
\usepackage{hyperref} 

\cftsetindents{section}{0em}{3em}   
\cftsetindents{subsection}{3.5em}{1.5em} 

\begin{document}

\title{
Maximum-precision charging of multi-qubit quantum batteries
}

\author{Davide Rinaldi\,\orcidlink{0009-0002-2562-0807}} 
\affiliation{Dipartimento di Fisica ``A. Volta'', Universit\`a di Pavia, via Bassi 6, 27100 Pavia, Italy}

\author{Radim Filip\,\orcidlink{0000-0003-4114-6068}}
\affiliation{Department of
Optics, Faculty of Science, Palacký University, 17. listopadu 12, 77900 Olomouc, Czech Republic}

\author{Dario Gerace\,\orcidlink{0000-0002-7442-125X}}
\affiliation{Dipartimento di Fisica ``A. Volta'', Universit\`a di Pavia, via Bassi 6, 27100 Pavia, Italy}

\author{Giacomo Guarnieri\,\orcidlink{0000-0002-4270-3738}}
\affiliation{Dipartimento di Fisica ``A. Volta'', Universit\`a di Pavia, via Bassi 6, 27100 Pavia, Italy}
\affiliation{INFN Sezione di Pavia, Via Agostino Bassi 6, I-27100 Pavia, Italy}

\begin{abstract}
{Precision, robustness, and efficiency are central requirements for quantum technologies. We show that genuine quantum features combined with non-Gaussianity enable the simultaneous optimization of these properties in a quantum battery–charging process. Using a generalized Jaynes–Cummings interaction as a paradigmatic light–matter interaction model, we apply the Full Counting Statistics to characterize stochastic energy exchanges between a stack of qubits and a single-mode bosonic field. We demonstrate that a sequential charging protocol driven by a non-Gaussian quantum field yields high performance in charging precision, which remains maximal even under suboptimal operating conditions. Our results establish the use of non-Gaussian quantum-states in battery charging as a robust route to a quantum precision advantage over protocols based on Gaussian states, achieved through the suppression of detrimental quantum fluctuations.
}

\end{abstract}

\maketitle

The role of energy in Physics is related to the capability of a system to perform work. An energy-based perspective can be effectively adopted to study quantum devices, ranging from quantum information technologies \cite{zaiser2016enhancing,degen2017quantum, Eisert1999,arute2019quantum,ladd2010quantum} to thermal engines and refrigerators~\cite{pekola2015towards,rossnagel2016single,maslennikov2019quantum,von2019spin,klaers2017squeezed,Onishchenko2024, mitchison2019refrigerators, cangemi2024quantum}. Recently, this viewpoint has fostered the emergence of a research field focused on the analysis of energy exchanges between quantum systems. The models developed in this context, where energy plays a central role, fall under the category of so-called quantum batteries~\cite{ferraro2026Perspective, campaioli2024colloquium, alicki2013, andolina2018charger,
CampaioliPRL2017, Barra2019, caravelli2020random, Tacchino2020, shaghaghi2023lossy, yang2024three, delmonte2021characterization, Barra1, Barra2, uwefisher_beneficial, campbell2025roadmap, morrone2023daemonic, grazi2024controlling_integrablespin, morroneTHESIS, medina2025anomalous_ergotropic_mpemba, sathe2025universally, campbell2025ergotopytransport, andolina2025genuine_anharmonic, andolina2019extractable, garcia2020fluctuations, tirone2023work, tirone2025quantum, lu2025topological, sarkar2025fluctuation, di2024local, cavaliere2025dynamical_blockade, hadipour2025amplified, hu2025enhancing_atomic, polo2025centrone, mitchison2021linear_feedback, ahmadi2024nonreciprocal, downing2024hyperbolic, malavazi2025charge, shastri2025dephasing, ferraro2018high,canzio2025single, camposeo2025quantum, hu2022optimal, elyasi2025experimental, 
elghaayda2025performance, razzoli2025cyclic, zhang2025single_ion, maillette2023experimental, joshi2022experimental, rinaldi2025reliable}. These models are so defined \cite{alicki2013} due to their classification of energy exchanges into three main processes: energy injection, energy storage, and on-demand energy delivery. This field is currently in the spotlight \cite{ferraro2026Perspective, campaioli2024colloquium}: numerous theoretical works on quantum batteries, aimed at exploring the fundamental limits of quantum thermodynamics \cite{grazi2024controlling_integrablespin, campbell2025roadmap, morrone2023daemonic, morroneTHESIS, medina2025anomalous_ergotropic_mpemba, sathe2025universally, campbell2025ergotopytransport, andolina2025genuine_anharmonic, andolina2019extractable, garcia2020fluctuations, tirone2023work, tirone2025quantum, lu2025topological, sarkar2025fluctuation, di2024local} and designing increasingly efficient and noise-resilient devices \cite{cavaliere2025dynamical_blockade, hadipour2025amplified, hu2025enhancing_atomic, polo2025centrone, mitchison2021linear_feedback, ahmadi2024nonreciprocal, downing2024hyperbolic, malavazi2025charge, shastri2025dephasing, ferraro2018high,canzio2025single}, are accompanied by experimental realizations \cite{camposeo2025quantum} on superconducting apparatuses \cite{hu2022optimal, elyasi2025experimental, elghaayda2025performance, razzoli2025cyclic}, trapped-ion platforms \cite{zhang2025single_ion}, quantum dots \cite{maillette2023experimental}, and nuclear-spin systems \cite{joshi2022experimental}. 

However, dealing with energy exchanges at a quantum level means that energy fluctuations cannot be neglected: their magnitude can assume values comparable to the average energy exchanges themselves. A simple (and yet effective) method to quantify energy fluctuations is the Full Counting Statistics~\cite{esposito2009nonequilibrium, landi2022nonequilibrium, landi2024current_fluctuations, brenes2023particle}, with which process-dependent thermodynamical quantities can be properly examined. We employ such technique to quantify the charging process in a Jaynes-Cummings \cite{stenholm1973quantum, smirne2010nakajima, bocanegra2024invariant, larson2021jaynes} quantum battery, i.e., a battery model consisting of a qubit (i.e., the battery) coupled to a single-mode harmonic oscillating field, typically confined in a resonating cavity (i.e., the charger), be it electromagnetic or mechanical. 

In this Letter, we leverage on the concepts and tools of Quantum Thermodynamics to actually implement realistic quantum batteries with proven advantages in maximizing the charging precision. We specifically demonstrate that a multi-qubit quantum battery can be operated at a low level of quantum fluctuations, crucial to envision a reliable quantum device, when a genuine quantum non-Gaussian state is employed as a charger. Such a quantum battery can then be employed to perform fine precision tasks, such as quantum state preparation~\cite{girolami2019difficult, wollack2022quantum, paraoanu2024high} or energy flow monitoring in quantum devices~\cite{graziadibello2025, dotsenko2020autonomous_maxwell_demon, franchi2022decoherence}.

\begin{figure}[t]
    \centering
    \includegraphics[width=0.45\textwidth]{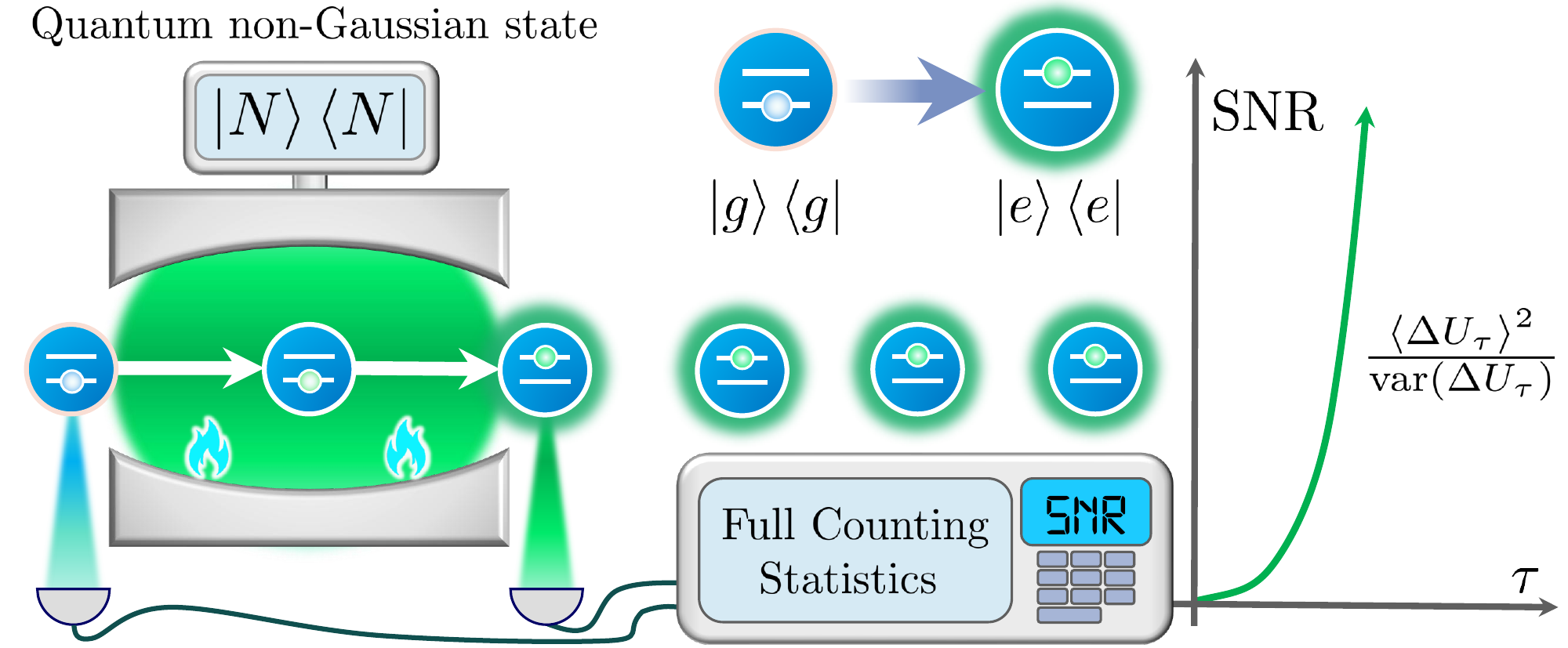}
    \caption{Multi-qubit JC quantum battery for precision-focused qubit charging with a Fock state cavity. The achieved charging advantage, quantified by the process-dependent SNR via the FCS, is both quantum non-Gaussian and operational, and it remains remarkably robust even beyond the ideal JC model and in realistic scenarios involving noise, dissipation, and parameter-tuning errors.    
    }
    \label{fig: FIG1}
\end{figure}

\textit{Model and methods}. -- The paradigmatic model for our quantum battery is based on the Jaynes-Cummings (JC) interaction \cite{stenholm1973quantum, smirne2010nakajima, bocanegra2024invariant, larson2021jaynes}. Drawing inspiration from microwave cavity QED experiments \cite{raimond2001manipulating}, this model describes a qubit with a resonance frequency $\omega_{\text{qub}}$, coupled to a single mode of a cavity harmonic oscillator (such as an electromagnetic or mechanical field operator), characterized by its frequency $\omega_{\text{cav}}$, as depicted in Fig.~\ref{fig: FIG1}. The qubit-cavity coupling is governed by the interaction strength $g$, which is assumed to remain constant over time. In the broadly-available weak-coupling regime, where $g/\omega_0 \ll 1$, with $\omega_{\text{qub}} \simeq \omega_0 \simeq \omega_{\text{cav}}$, the counter-rotating terms in the Rabi Hamiltonian can be neglected by applying the rotating wave approximation (RWA) \cite{stenholm1973quantum}: we quantify the error made by performing the RWA in the Supplemental Material (SM) \cite{suppmat}.
To extend this framework to $M$ non-interacting qubits coupled to the same harmonic oscillator, we introduce the following general Hamiltonian:
\begin{equation}
    \label{eq: GENERAL Hamiltonian}
        \hat{H}(t) = \hat{H}_{\text{free}} +  \sum_{j=1}^{M}  \hbar g A_j(t)(\hat{\sigma}^{(j)}_+ \hat{a} + \hat{\sigma}^{(j)}_- \hat{a}^\dagger) \, .
\end{equation}
When the factor $A_j(t) = 0$ $\forall j\neq1$ and $A_1(t) = 1$ $\forall t$, we recover the standard JC model \cite{stenholm1973quantum, Wallraff2004,Hennessy2007,DeLiberato2009,schuster2007resolving,forn2019ultrastrong, beaudoin2011dissipation}, which can be analytically solved by computing the evolution operator $\hat{\mathcal{U}}(t, 0) = \mathcal{T}e^{-\frac{i}{\hbar}\int_0^t \hat{H}(t')dt'}$ \cite{stenholm1973quantum, smirne2010nakajima, bocanegra2024invariant, larson2021jaynes}. Here, $\hat{H}_{\text{free}} = \sum_{j=1}^M \hat{H}^{(j)}_{\text{qub}} + \hat{H}_{\text{cav}}$, $\hat{H}^{(j)}_{\text{qub}} = \hbar\omega_{\text{qub}} \frac{1}{2}\hat{\sigma}_z^{(j)}$, and $\hat{H}_{\text{cav}} = \hbar \omega_{\text{cav}}\hat{a}^\dagger\hat{a}$, in which $\hat{\sigma}^{(j)}_{x,y,z}$ denote the standard Pauli matrices for the $j$-th qubit, $\hat{\sigma}^{(j)}_{\pm} = \frac{1}{2}( \hat{\sigma}^{(j)}_{x}\pm i\hat{\sigma}^{(j)}_{y})$, and $\hat{a}$ ($\hat{a}^\dagger$) is the annihilation (creation) operator of a single excitation in the harmonic single-mode field.
The $j$-th qubit state basis $\{\ket{e}_j, \ket{g}_j\}$, eigenvectors of $\hat{\sigma}^{(j)}_z$, defines the pure excited and ground state, respectively. 
We define the $j$-th qubit state at time $t$ as $\hat{\rho}^{(j)}_{\text{qub}}(t)$, and the cavity mode state as $\hat{\rho}_{\text{cav}}(t)$; in addition, at the initial time $t=0$, the qubits and cavity mode states are assumed to be factorized: $\hat{\rho}_{\text{total}}(0) = \bigotimes_{j=1}^{M}\hat{\rho}^{(j)}_{\text{qub}}(0)\otimes\hat{\rho}_{\text{cav}}(0)$. The cavity field state is then initialized \cite{dotsenko_haroche_2008reconstruction} in either a $N$-photon Fock state, $\ket{N}\bra{N}$, or a coherent state $\ket{\alpha}\bra{\alpha}$, for comparison, in which 
$|\alpha|^2$ is the average number of photons, or also in a quantum Gaussian state, e.g., a squeezed coherent state $\ket{\tilde{\alpha},\zeta}\bra{\tilde{\alpha},\zeta}$, with $\zeta$ being the squeezing parameter \cite{adesso2014continuous}. The charging process consists of using the Hamiltonian interaction (\ref{eq: GENERAL Hamiltonian}) with the charger to induce ground-to-excited state transitions in each qubit. The goal is to do so with the maximum precision (i.e., lowest variance) possible on the energy exchanged between the cavity and the qubit \textit{during a time window} $\tau$.  

With the aim of quantitatively assessing the behavior of this energy exchange, and thus characterizing its fluctuations over the interaction time, we focus on the stochastic variable $\Delta U_\tau$ \cite{hickey2013time}. We refer to this variable as the ``qubit internal energy'', as it is directly related to the measurement outcomes of a specific observable: the Hamiltonian $\hat{H}_{\text{qub}} = \hbar\omega_{\text{qub}}\frac{1}{2}\hat{\sigma}_z$, evaluated at two distinct times.
Here we use theoretical techniques falling within the \textit{Full Counting Statistics} (FCS) \cite{esposito2009nonequilibrium}. Within FCS, the statistics of $\Delta U_\tau$ are fully described by a qubit-specific generating function $\mathcal{G}_\tau(\chi)$, such that $\langle\Delta U_\tau^n \rangle = (-i)^n \frac{\partial^n\mathcal{G}(\chi)}{\partial\chi^n}\bigl\rvert_{\chi = 0}$. $\mathcal{G}_\tau(\chi)$ can be experimentally measured \cite{goold2014measuring_gen_func} and compared to its theoretically derived form, whenever the underlying dynamical model is known. Specifically, the generating function is computed as: $\mathcal{G}_\tau(\chi) = \text{Tr}\bigl\{ \hat{\rho}_{\chi}(t) \bigr\}$, where $\hat{\rho}_{\chi}(t) = \hat{\mathcal{U}}_{\frac{\chi}{2}}(\tau,0) \hat{\rho}_{\text{JC}}(0) \hat{\mathcal{U}}^\dagger_{-\frac{\chi}{2}}(\tau,0)$ and $\hat{\mathcal{U}}_{\frac{\chi}{2}}(\tau,0) = e^{i\frac{\chi}{2}\hat{H}_{\text{qub}} }\hat{\mathcal{U}}(\tau,0)e^{-i\frac{\chi}{2}\hat{H}_{\text{qub}}}$ \cite{landi2022nonequilibrium, esposito2009nonequilibrium}.  The central figure of merit for our discussion is the Signal-to-Noise Ratio (SNR), which can be extracted from the statistics of $\Delta U_\tau$, i.e., formally defined as
\begin{equation}
    \label{eq: SNR formula}
    \text{SNR}(\Delta U_\tau) = \frac{\langle \Delta U_\tau\rangle^2}{\text{var}(\Delta U_\tau)}
\end{equation}
where $\text{var}(\Delta U_\tau) = \langle \Delta U^2_\tau \rangle - \langle \Delta U_\tau\rangle^2$. As it is evident from Eq.~(\ref{eq: SNR formula}), the SNR quantifies the signal amplitude (i.e., the mean exchanged energy) as compared to the magnitude of its fluctuation (i.e., its variance), and it therefore constitutes the best measure of the process precision. Within FCS, these \textit{process}-dependent quantities allow to better capture the complexity of the fluctuating energy exchanges. This represents a remarkable conceptual leap from standard techniques, which employ \textit{state}-dependent figures of merit focused only on the system state at a given time, although the latter have been applied in other thorough analyses, including Refs.~\cite{polo2025centrone, cavaliere2025dynamical_blockade, canzio2025single}. For a comparison, we then compute the fidelity between the qubit state, $\hat{\rho}_{\text{qub}}(t)$, and the target state, $\hat{\rho}_{\text{target}} = \ket{e}\bra{e}$, defined as 
\begin{equation}
    \label{eq: Fidelity}
    F(t) = \text{Tr}\bigl\{ \bigl[ \hat{\rho}^{\frac{1}{2}}_{\text{qub}}(t) \hat{\rho}_{\text{target}} \hat{\rho}^{\frac{1}{2}}_{\text{qub}}(t)  \bigr]^{\frac{1}{2}} \bigr\} ^2 
\end{equation}
In the following, we show two different protocols allowing to charge more than a single qubit at a time, by using the harmonic oscillator state as a charger, with the aim of achieving the maximal charging precision: (i) sequential and (ii) parallel protocols, respectively.

\begin{figure}[t]
    \centering
    \includegraphics[width=0.48\textwidth]{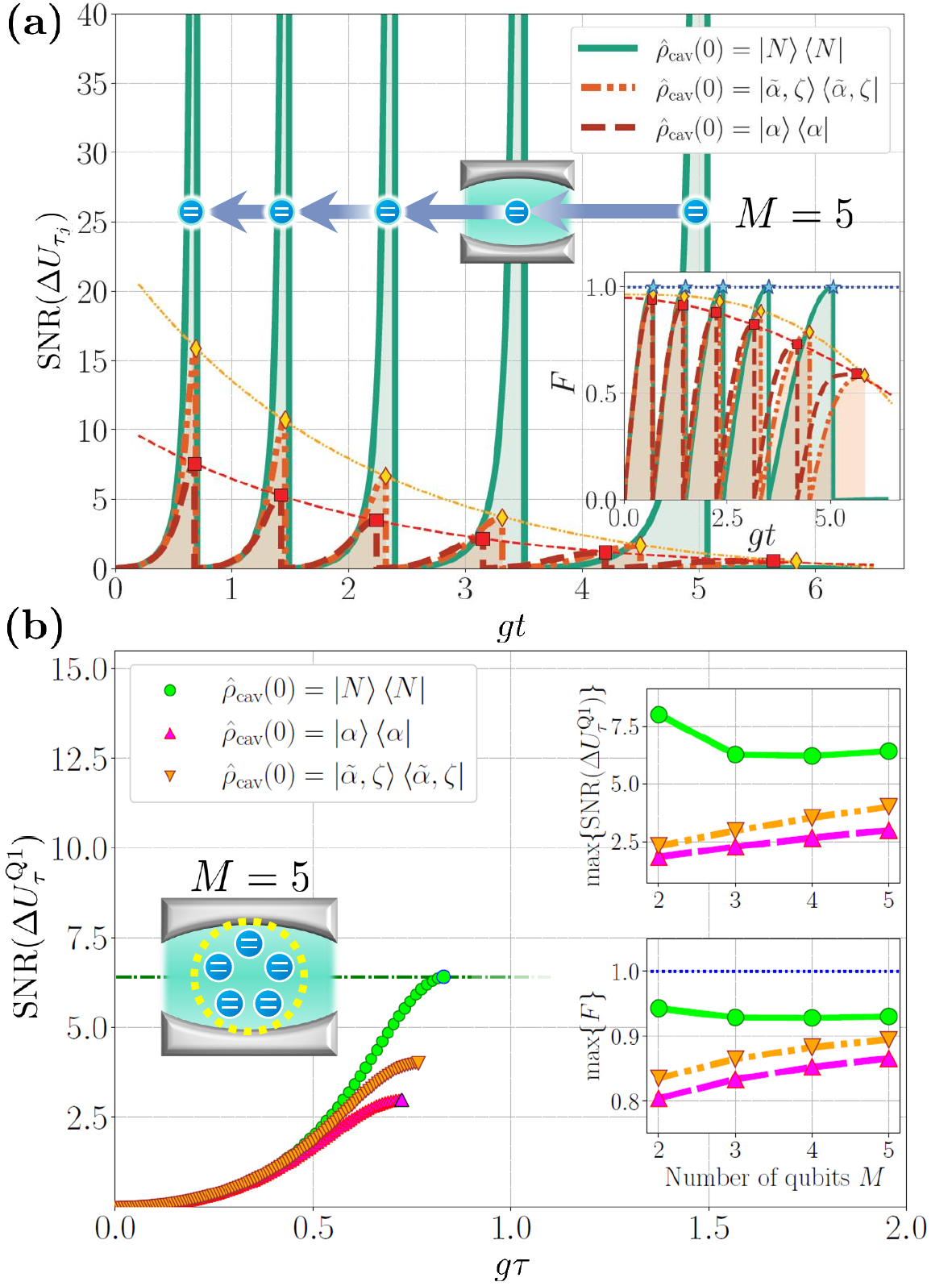}
    \caption{(a) Sequential charging of $M=5$ qubits, where $\text{SNR}(\Delta U_{\tau_i})$ is calculated for each qubit charge. The fidelity $F(t)$ between $\hat{\rho}^{(j)}_{\text{qub}}(t)$ and $\hat{\rho}^{(j)}_{\text{target}} = \ket{e}_j\bra{e}_j$ is depicted in the inset. While in the presence of idealized conditions and perfect tuning of the interaction time $\tau$ the Fock state SNR peaks can be arbitrarily high, the ones related to the other states decrease exponentially in time. The fidelity reaches $F=1$ only for the Fock state protocol; for the others, $F<1$ and the peaks decrease in time as $F\propto-(gt)^\gamma$, with $\gamma>0$. (b) Parallel charging of $M=5$ qubits. The $\text{SNR}(\Delta U_\tau^{\text{Q1}})$ is related to the sole qubit 1, and it is obtained by numerically solving the TC model. Upper inset: maximum SNR for $M\in\{2,3,4,5\}$ qubits. Lower inset: maximum fidelity $F$ between $\hat{\rho}^{(1)}_{\text{qub}}(t)$ and $\hat{\rho}^{(1)}_{\text{target}} = \ket{e}_1\bra{e}_1$, varying the number of qubits. Simulation parameters: $\langle n \rangle = N=5$, $\alpha = \sqrt{5}$ for the coherent state, $\zeta = 0.6$ and $\tilde{\alpha} \simeq 3.905$ for the squeezed coherent state, $g/\omega_{\text{qub}} = 10^{-2}$, and $\Delta\omega = 0$.} 
    \label{fig: FIG4}
\end{figure}

\textit{Sequential charging protocol.} -- Let us consider Eq.~(\ref{eq: GENERAL Hamiltonian}) with $A_j(t) = 1$ if $t\in[t_{j-1},t_{j})$, and $A_j(t)=0$ otherwise. Now, the $j$-th qubit interacts with the cavity during a time window $\tau_j = t_{j}-t_{j-1}$ (in general, $\tau_j \neq \tau_k$ for $j\neq k$, and for convenience $t_0 \equiv 0$). After the interaction, the $i$-th qubit is immediately decoupled from the cavity, and at the same time the $(j+1)$-th qubit is coupled to the oscillator, analogously to the mechanism that occurs within a collisional-model description \cite{ciccarello2022quantum}. \\
Due to the peculiarity of the model in Eq.~(\ref{eq: GENERAL Hamiltonian}), the Hamiltonians of different qubits commute with each other. This allows us to apply the same formalism of a preparatory work \cite{rinaldi2025reliable}, where the elementary case of a single qubit was discussed, to derive the SNR of the exchanged energy in a modular fashion, i.e., within each time window $\tau_j$. Importantly, we now keep track of the cavity state after each interaction: the $j$-th charging round begins with the $j$-th qubit in a fresh ground state, while the cavity state is modified to account for the qubit measurement-induced back-action \cite{suppmat} after the interaction with the $(j-1)$-th qubit. \\
As we show in Fig.~\ref{fig: FIG4}(a), the Fock state SNR on the energy $\Delta U_\tau$ injected in each qubit can be arbitrarily increased, thus being higher than the SNR obtained via coherent or Gaussian states, provided that the model parameters can be tuned to achieve the desired working conditions, such as perfect resonance and the absence of external disturbances. For that reason, in an idealized situation it is possible to perfectly charge $M$ qubits by initializing the cavity in a $N$-photon Fock state with $N=M$, a result confirmed by the fact that the fidelity between the $j$-th qubit states $\hat{\rho}^{(j)}_{\text{qub}}(t)$ and $\hat{\rho}^{(j)}_{\text{target}} = \ket{e}_j\bra{e}_j$ can always reach $F=1$ for each charging round, contrary to what happens if the cavity is prepared in a classical or a Gaussian state. Furthermore, other realistic scenarios will be discussed in what follows, showing that the quantum advantage persists even if the variance does not completely vanish. 

\textit{Parallel charging protocol.} -- Next, we analyze the parallel charging protocol, for which we resort to the Tavis-Cummings (TC) model \cite{fujii2004TWOatoms}, which describes multiple mutually non-interacting qubits, each simultaneously coupled to the single mode field confined to the cavity. This model cannot be formulated as a sequence of excitation exchanges: it thus provides insights into the nature of the Fock precision advantage, which can not be reduced to a mere swap of well-defined Fock states. The $M$-qubit TC Hamiltonian can be derived from Eq.~(\ref{eq: GENERAL Hamiltonian}) by setting $A_j(t) = 1$ $\forall j,t$. Such a model can be analytically solved for $M\in\{1, ..., 4\}$ qubits under the perfect resonance condition, namely $\Delta\omega = 0$ \cite{fujii2004TWOatoms, lu2025singleMISSINGTERM, fujii2004THREEFOURatoms}. We analytically study the case for $M=2$ as a proof of concept, and numerically solve the cases for $M\geq3$.

Here, we primarily considered the energy exchange of a single qubit, as well as the fidelity between the single qubit states $\hat{\rho}^{(j)}_{\text{qub}}(t)$ and $\hat{\rho}^{(j)}_{\text{target}} = \ket{e}_j\bra{e}_j$. Since the model is symmetric with respect to any permutation of the qubits, this is equivalent to study the fidelity between the $M$-qubit collective states $\hat{\rho}^{(1,...,M)}_{\text{qub}}(t) = \bigotimes_{j=1}^M \hat{\rho}_{\text{qub}}^{(j)}(t)$ and $\hat{\rho}^{(1,...,M)}_{\text{target}} = \bigotimes_{j=1}^M \ket{e}_j\bra{e}_j$. 

The statistics related to the energetics of the single qubit can be derived by rotating the evolution operator as $e^{i\frac{\chi}{2}\hat{\mathcal{O}} }\hat{\mathcal{U}}(\tau,0)e^{-i\frac{\chi}{2}\hat{\mathcal{O}}}$ with $\hat{\mathcal{O}} = \hat{H}_{\text{qub}}^{(1)} = \hbar\omega_{\text{qub}} \frac{1}{2}\hat{\sigma}_z^{(1)}$ (further details are provided in SM \cite{suppmat}). 
On the numerical side, an effective way to numerically calculate the qubits energy statistics is to solve a differential equation for the transformed density matrix $\hat{\rho}_\chi(t)$ of the total system \cite{landi2022nonequilibrium}:
\begin{equation}
\label{eq: Differential equation for rho_chi, Tavis-Cummings}
    \frac{d}{dt}\hat{\rho}_{\chi}(t) = -\frac{i}{\hbar}\bigl[ \hat{H}_{\text{TC},\chi} \hat{\rho}_{\chi}(t) - \hat{\rho}_{\chi}(t) \hat{H}_{\text{TC}, -\chi} \bigr]\\
\end{equation}
where $\hat{H}_{\text{TC}, \chi}$ is the modified TC Hamiltonian $\hat{H}_{\text{TC}, \chi} = \hat{H}_{\text{free}} + \hat{H}^{(1)}_{\text{int}, \chi}$, with $\hat{H}^{(1)}_{\text{int}, \chi} = e^{i\frac{\chi}{2}\hat{H}_{\text{qub}}^{(1)}} \hat{H}_{\text{int}} e^{-i\frac{\chi}{2}\hat{H}_{\text{qub}}^{(1)}}$, being $ \hat{H}_{\text{int}} = \hbar g \sum_{j=1}^M (\hat{\sigma}_+^{(j)}\hat{a} + \hat{\sigma}_-^{(j)}\hat{a}^\dagger) $. As reported in Fig.~\ref{fig: FIG4}(b) for the parallel charging of $M=5$ qubits within a single time window $\tau$, the advantage in the SNR offered by the Fock state cannot be arbitrarily enhanced by optimizing the model parameters, unlike what occurs with the sequential charging protocol. As well, the charging of a single qubit can not be perfectly performed, since $F<1$ even for the Fock state protocol. Furthermore, as $M$ increases, the performance of the Fock state cavity becomes increasingly comparable to that of the coherent and Gaussian states. For these reasons, we conclude that the perfect charging of a stack of qubits with a single-mode field can only be achieved by employing both a Fock state reservoir and a sequential protocol.

\textit{Realistic scenarios.} -- Two questions arise: Is the sequential protocol robust in realistic scenarios? If so, what conditions maximize its precision described by the process-dependent SNR~(\ref{eq: SNR formula})? To address this question, we investigate a broad variety of situations, accounting for the effects of noise, dissipation, and non-ideal parameter choices. We also relax the JC model by introducing several kinds of generalizations. These results are thoroughly presented in the Supplemental Material \cite{suppmat, leonhardt2003quantum, chen2014qubit_FAST_TUNABLE_COUPLING, stehlik2021tunable, plenio2005mutual_info, khanahmadi2023multimode_FOR_DECAY_RATE, braak2019symmetries, xie2014anisotropicRabi, nielsen_chuang}. Here, as a paradigmatic example of a source of non-ideality, we discuss the impact of thermal photons, which can degrade the quantum properties of the cavity state. We do so by simulating a sequential charging of $M=5$ qubits. For each charging process of the entire stack of qubits, we set the mean number of initial photons $\langle n \rangle$ in the cavity equal to a value among $\{1,2,3,4,5\}$. We then simulate the battery charging process for both Fock and Gaussian initial cavity states. 
In all cases, we consider the worst-case scenario of a noisy Fock cavity state compared to an ideal, noiseless Gaussian state.
To do so, we introduce the following quantity, parametrized by the number of thermal photons, $\bar{n}_{\text{th}}$:
$\overline{\text{SNR}}_{\bar{n}_{\text{th}}, \langle n \rangle} \equiv \frac{1}{\langle n \rangle} \sum_{j=1}^5 \max_{\tau_j}\{\text{SNR}_j(\bar{n}_{\text{th}}, \langle n \rangle)\}$. It represents an averaged SNR, normalized with respect to the field energy. This ensures the fairness of our findings, as their accuracy is no longer limited by the resources (i.e., the initial number of photons) available to the experimenter. The figure of merit we use is defined as the difference between the quantity calculated for both the Fock and Gaussian cavity states.:
\begin{equation}
    \label{eq: Figure of merit for noise analysis}
    D_{\bar{n}_{\text{th}}, \langle n \rangle} = \overline{\text{SNR}}_{\bar{n}_{\text{th}}, \langle n \rangle}\biggr\rvert_{\text{Fock}} - \overline{\text{SNR}}_{\bar{n}_{\text{th}}, \langle n \rangle}\biggr\rvert_{\text{Gaussian}}
\end{equation}
From the definition of $D_{\bar{n}_{\text{th}}, \langle n \rangle}$, it is evident that a precision advantage of the Fock state over the Gaussian state corresponds to the condition $D_{\bar{n}_{\text{th}}, \langle n \rangle} > 0$. Moreover, $D_{\bar{n}_{\text{th}}, \langle n \rangle}$ also accounts for situations in which a cavity state with $\langle n \rangle$ initial photons does not perfectly charge the first $n$ qubits, but delivers the remaining energy to the subsequent ones. 

\begin{figure}[t]
    \centering
    \includegraphics[width=0.48\textwidth]{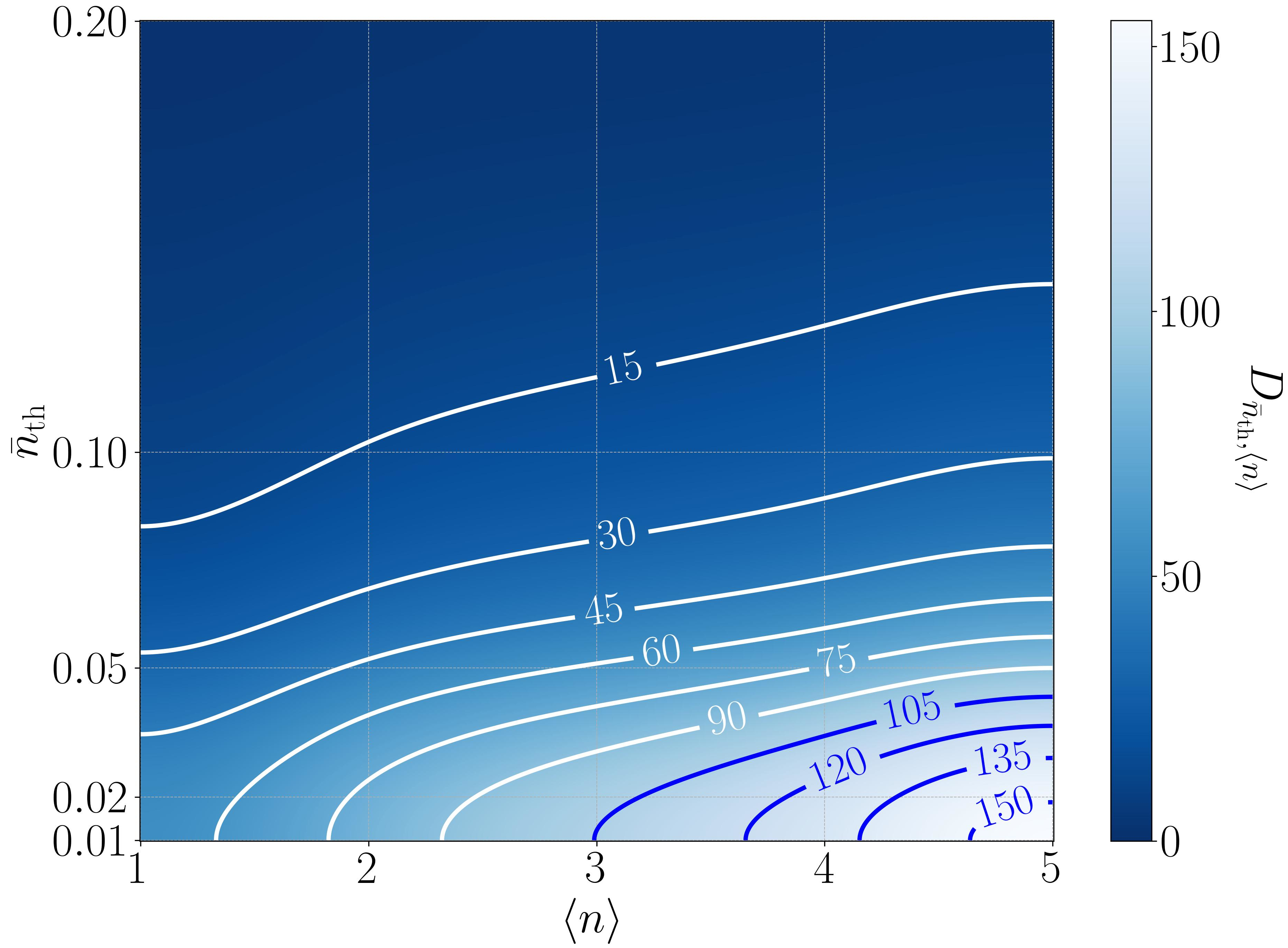}
    \caption{Effect of the presence of $\bar{n}_{\text{th}}$ thermal photons on the difference $D_{\bar{n}_{\text{th}},\langle n \rangle}$. To compare with phase-insensitive Fock states, we used a Gaussian coherent squeezed state $\ket{\zeta, \tilde{\alpha}}\bra{\zeta, \tilde{\alpha}}$ with randomized phases \cite{pradana2019quantum, suppmat}. Simulation parameters: $N = \langle n \rangle$ for the Fock state; $\zeta$ and $\tilde{\alpha}$ are optimized to give the best performances for the squeezed coherent state; $g/\omega_{\text{qub}} = 10^{-2}$, and $\Delta\omega = 0$.} 
    
    \label{fig: FIG7}
\end{figure}

To model the presence of thermal photons in the Fock state, we follow Ref.~\cite{ritboon2022sequential}, where a $N$-photon Fock state exposed to a source of additive thermal noise for a short time is expressed in terms of the map $\mathcal{K}$ such that $\hat{\rho}_{\text{Fock}}^{\text{Th}} = \mathcal{K}\bigl(\ket{N}\bra{N}\bigr) =\sum_{m=0}^{\infty} p_{m}(\bar{n}_{\text{th}}) \ket{m}\bra{m}$ (details are reported in SM \cite{suppmat}). From Fig.~\ref{fig: FIG7}, we observe that the difference $D_{\bar{n}_{\text{th}},\langle n \rangle}$ remains positive even for $\bar{n}_{\text{th}} = 0.2$ on average; for lower values, such as $\bar{n}_{\text{th}} = 0.02$, $D_{\bar{n}_{\text{th}},\langle n \rangle}$ can become significantly large, ranging from $D_{\bar{n}_{\text{th}},\langle n \rangle=1}\sim50$ to $D_{\bar{n}_{\text{th}},\langle n \rangle=5}\sim150$. Therefore, the precision advantage of the Fock state charging protocol proves robust against thermal noise, as long as the noise level remains below a critical threshold. Notably, the performance of the non-Gaussian quantum protocol improves with the mean photon number $\langle n \rangle$; furthermore, the tolerance to noise scales with the initial photon number $N$ stored in the Fock state, reinforcing the protocol’s resilience at higher photon populations.


\textit{Conclusions.} -- 
We demonstrated that a multi-qubit Jaynes-Cummings (JC) quantum battery can achieve maximally precise charging, i.e., high precision performances even in presence of non-ideal working conditions, by using an optimized protocol and harnessing non-Gaussian quantum states. We employed the Full Counting Statistics (FCS) to compute the process-dependent Signal-to-Noise Ratio (SNR) associated with the energy injected into the battery, and compared  sequential and parallel charging protocols of $M$ qubits. Our results show that under ideal, noiseless conditions, a cavity prepared in a $N$-photon Fock state can perfectly charge $M = N$ qubits sequentially, whereas parallel charging fails to achieve perfect charging.
We then focused on the sequential protocol, evaluating its robustness in realistic scenarios, here exemplified by the detrimental effects of spurious thermal photons. A more complete analysis and thorough discussions are developed in SM \cite{suppmat}. Despite the negative impact of the considered external limiting factors, we demonstrate that the maximum charging precision is achievable when using a genuine non-Gaussian quantum state for sequential qubit charging. The proposed protocol, albeit simple, is therefore proven to be resilient in realistic working conditions, thus being feasible for direct implementations. Such a result serves as a clear signature of the battery quantum nature, and suggests that this effect should be experimentally observable: it applies indeed to recent experiments in circuit QED \cite{hofheinz2008generation}, trapped ions \cite{monroe_kim2021programmable, podhora2022quantum, podhoraTHESIS}, and quantum electromechanics \cite{rahman2025genuine}.

\textit{Acknowledgments.} -- R.F. acknowledges Grant No. 21-13265X of the Czech Science Foundation. D.G. acknowledges the ``National Quantum Science Technology Institute'' (NQSTI) within the PNRR Project No. PE0000023. G.G. acknowledges support from the Ministero
dell’Università e della Ricerca (MUR) under the ``Rita Levi-Montalcini'' grant and from INFN. D.R. acknowledges A. Viola, E. Tumbiolo, G. Candreva, A. E. Mazzarone, L. Frau, D. Tona, G. Coppola, G. F. Conte, F. M. Quetti, L. Marti, and R. G. Pozzi for the profound scientific discussions. \\

\textit{Data Availability.} -- The data that support the findings of this article are openly available \cite{data_availability}.


\bibliographystyle{apsrev4-1}

%


\clearpage

\onecolumngrid

{\Large\centering\textbf{Supplemental Material}} \\

\renewcommand{\theequation}{S.\arabic{equation}}

\setcounter{equation}{0}

\renewcommand\thefigure{S.\arabic{figure}} 

\setcounter{figure}{0}


\par\noindent
In the present Supplemental Material, we offer additional details about the results presented in the main text. Below is a list of the contents of this additional material, serving as a guide for the interested reader.

\tableofcontents

\clearpage
\section{Methodology}
\label{app: Methodology}
In that section, we clarify the method we employed to apply the Full Counting Statistics (FCS) to monitor the charging process of a stack of qubits by computing the signal-to-noise ratio (SNR) of the energy injected into the qubits. 

\subsection{Full Counting Statistics for battery characterization}
\label{subsec: Application of the FCS}
We focus specifically on the sequential protocol, which entails certain subtleties worth emphasizing and motivating, as the FCS application here involves multiple measurement steps. In contrast, the parallel protocol is conceptually simpler to analyze, requiring only two measurements, one at the beginning and one at the end, as prescribed by the FCS. For the sequential protocol, however, the methodology we adopt in the main text and in this Supplemental Material can be seen as a battery \textit{characterization}, with each of the $M$ qubits measured twice, resulting in a total of $2M$ measurements by the protocol's end. While such measurements could, in principle, affect the protocol's behavior, we demonstrate below (see Sect.~\ref{subsec: Actual battery charging}) that this influence is minimal and practically negligible.

We now turn to the characterization strategy for the sequential protocol. The procedure can be outlined as follows: \\
\begin{figure}[htp!]
    \centering
\includegraphics[width=0.9\linewidth]{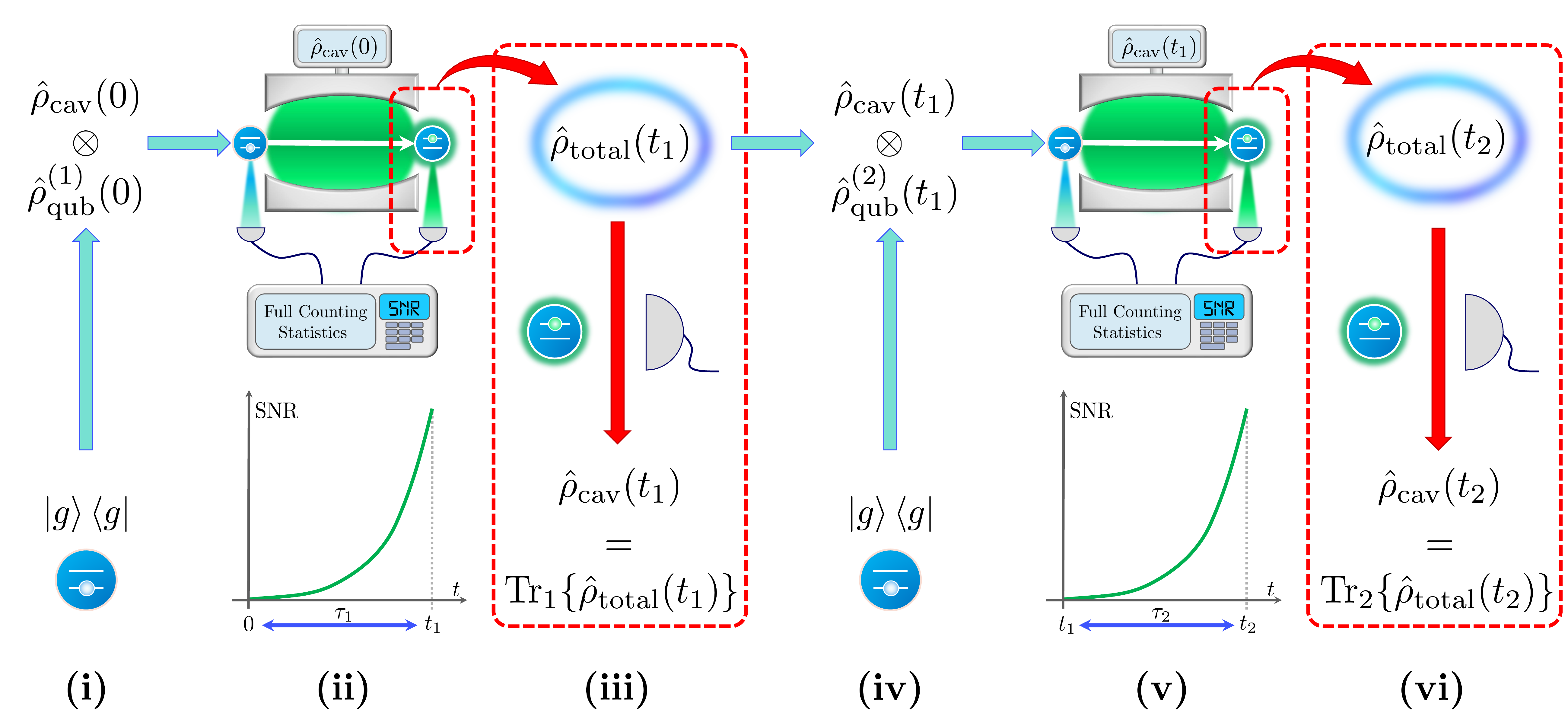}
    \caption{Application of the FCS to characterize the battery SNR in the sequential protocol. The parallel protocol is conceptually simpler, as it involves two measurements only (at the beginning and at the end of the collective charging process).}
    \label{fig: Methodology scheme}
\end{figure}
\begin{enumerate}[label=(\roman*)]
    \item At $t=0$, the total joint state $\hat{\rho}_{\text{total}}(0)$ is initialized as a tensor product of the first qubit's state, $\hat{\rho}^{(1)}_{\text{qub}}(0) = \ket{g}\bra{g}$, and the initial cavity state $\hat{\rho}_{\text{cav}}(0)$.
    \item The charging of the first qubit proceeds from $t=0$ to $t=t_1$ (with duration $\tau_1 = t_1$), during which the qubit interacts with the cavity. We monitor the process by imagining that two projective energy measurements are performed at $t=0$ and $t=t_1$, respectively. Then, the FCS enables the computation of the statistics of $\Delta U_{\tau_1}$, defined as the difference between measurement outcomes, from which we derive $\text{SNR}(\Delta U_{\tau_1})$.
    \item We obtain the final state after the interaction by solving the associated Schr\"odinger equation (or, if dissipation is included, by solving the appropriate master equation). Thus, without measurements, the joint state $\hat{\rho}_{\text{total}}(t_1)$ at the end of the charging process may, in general, not be factorized. However, within the FCS framework, the second energy measurement projects the qubit into a pure state: $\ket{e}\bra{e}$ or $\ket{g}\bra{g}$. Consequently, the total final state collapses into a tensor product of the qubit state and the cavity state. To remain consistent with this picture, we compute the reduced cavity state $\hat{\rho}_{\text{cav}}(t_1)$ by tracing over the qubit's degrees of freedom, thereby neglecting any emergent correlations. We further justify this approach below.
    \item At $t=t_1$, the new joint state is constructed as a tensor product of the second qubit's ground state, $\hat{\rho}^{(2)}_{\text{qub}}(t_1) = \ket{g}\bra{g}$, and $\hat{\rho}_{\text{cav}}(t_1)$.
    \item The charging of the second qubit is monitored analogously via the FCS: two energy measurements are performed at $t=t_1$ and $t=t_2$ (with duration $\tau_2 = t_2 - t_1$), and the SNR of the energy difference $\Delta U_{\tau_2}$ is computed.
    \item The cavity state at $t=t_2$, $\hat{\rho}_{\text{cav}}(t_2)$, is again obtained by tracing over the qubit's degrees of freedom. This state is used to construct the total joint state at $t=t_2$ as a tensor product of the third qubit's ground state, $\hat{\rho}^{(3)}_{\text{qub}}(t_2) = \ket{g}\bra{g}$, and $\hat{\rho}_{\text{cav}}(t_2)$. The process then repeats iteratively.
\end{enumerate}

\subsection{Measurement back-action}
\label{subsec: Measurement back-action}
In step (iii), to construct the joint state of the \textit{second} qubit and the cavity after the \textit{first} qubit's charging process, we compute the tensor product $\hat{\rho}_{\text{qub}}^{(2)}(t_1)\otimes \hat{\rho}_{\text{cav}}(t_1)$, where $\hat{\rho}_{\text{qub}}^{(2)}(t_1) = \ket{g}\bra{g}$. The cavity state $\hat{\rho}_{\text{cav}}(t_1)$ is derived by tracing the total joint state $\hat{\rho}_{\text{total}}(t_1)$ over the \textit{first} qubit's degrees of freedom:
\begin{equation}
    \label{eq: Cavity state after the first interaction}
    \hat{\rho}_{\text{cav}}(t_1) = \text{Tr}_1\{ \hat{\rho}_{\text{total}}(t_1) \}.
\end{equation}
This procedure accounts for any back-actions on the cavity state due to the measurement process. Indeed, if the qubit and the cavity are entangled, a projective measurement on the qubit influences the cavity state as well. Consequently, the charging protocol could be affected by the perturbation introduced by the FCS measurement requirements, and its implications must be considered.

The effects of a projective measurement on a state $\hat{\rho}$ are modeled in a standard way \cite{nielsen_chuang} by introducing a set of measurement operators $\hat{P}_m$ such that $\sum_m \hat{P}_m^\dagger\hat{P}_m = \hat{\mathds{I}}$, each associated with the measurement result $\varepsilon_m$. After a measurement described by these operators, the system's state becomes:
\begin{equation}
    \label{eq: State after measurement, general}
    \hat{\rho}' = \sum_m p_m \hat{\rho}_m = \sum_m \hat{P}_m \hat{\rho} \hat{P}_m^\dagger,
\end{equation}
where:
\begin{equation}
    \label{eq: Probability pm and renormalized state}
    p_m = \text{Tr}\{\hat{P}_m \hat{\rho} \hat{P}_m^\dagger\}, \quad \hat{\rho}_m = \frac{\hat{P}_m \hat{\rho} \hat{P}_m^\dagger}{p_m}
\end{equation}
are the probability $p_m$ of obtaining the measurement result $\varepsilon_m$ and the associated collapsed state $\hat{\rho}_m$, respectively. We now apply this formalism to the two-qubit charging process described in Fig.~\ref{fig: Methodology scheme}. The overall state of the two qubits and the cavity at $t=0$ is:
\begin{equation}
    \label{eq: Initial state, two qubits and cavity}
    \hat{\rho}_{\text{total}}(0) = \ket{g}_1\bra{g}_1\otimes\ket{g}_2\bra{g}_2\otimes\hat{\rho}_{\text{cav}}(0).
\end{equation}
The FCS involves two measurements of the qubit energy. Consequently, the set of measurement operators corresponding to the energy outcomes $\varepsilon_e = +\hbar\omega_\text{qub}/2$ and $\varepsilon_g = -\hbar\omega_\text{qub}/2$ is given by:
\begin{equation}
    \label{eq: Measurement operators for the qubit}
    \hat{P}_e^{(j)} = \ket{e}_j\bra{e}_j, \quad \hat{P}_g^{(j)} = \ket{g}_j\bra{g}_j,
\end{equation}
where $j=1,2$ distinguishes between the first and second qubit. The first measurement at $t=0$ leaves the state unchanged, as the first qubit is in the state $\ket{g}_1\bra{g}_1$, factorized from the state of the other subsystems. After the first interaction from $t=0$ to $t=t_1$, the overall state, denoted $\hat{\rho}_{\text{total}}(t_1)$, may in general be entangled. At this stage, we perform the second measurement, updating the state according to (\ref{eq: State after measurement, general}):
\begin{equation}
    \label{eq: Overall state after measurement of 1st qubit}
    \hat{\rho}'_{\text{total}}(t_1) = \ket{e}_1\bra{e}_1 \hat{\rho}_{\text{total}}(t_1) \ket{e}_1\bra{e}_1 + \ket{g}_1\bra{g}_1 \hat{\rho}_{\text{total}}(t_1) \ket{g}_1\bra{g}_1.
\end{equation}
Notably, the first qubit's state becomes factorized from the second qubit and the cavity after this measurement. For example, if the measurement yields $\varepsilon_e = +\hbar\omega_\text{qub}/2$, the collapsed state is:
\begin{equation}
    \label{eq: Collapsed state |e> after measurement of 1st qubit}
    \hat{\rho}^{(e)}_{\text{total}}(t_1) = \frac{\ket{e}_1\bra{e}_1 \hat{\rho}_{\text{total}}(t_1) \ket{e}_1\bra{e}_1}{p_e} = \ket{e}_1\bra{e}_1 \otimes \frac{\bra{e}_1 \hat{\rho}_{\text{total}}(t_1) \ket{e}_1}{p_e}.
\end{equation}
Equation~(\ref{eq: Collapsed state |e> after measurement of 1st qubit}) clearly shows that the second FCS measurement completely destroys any qubit-cavity correlations. Besides, since the first qubit's state remains factorized from the state of the other subsystems, the partial trace of $\hat{\rho}'_{\text{total}}(t_1)$ over the first qubit yields:
\begin{equation}
    \label{eq: Partial trace on qubit 1}
    \text{Tr}_1\{\hat{\rho}'_{\text{total}}(t_1)\} = \bra{e}_1 \hat{\rho}_{\text{total}}(t_1) \ket{e}_1 + \bra{g}_1 \hat{\rho}_{\text{total}}(t_1) \ket{g}_1 = \text{Tr}_1\{\hat{\rho}_{\text{total}}(t_1)\}.
\end{equation}
In other words, computing the cavity state as $\hat{\rho}_{\text{cav}}(t_1) = \text{Tr}_2\bigl\{\text{Tr}_1\{ \hat{\rho}_{\text{total}}(t_1) \}\bigr\}$
is equivalent to deriving it via (\ref{eq: Cavity state after the first interaction}), since the second qubit's state, $\ket{g}_2\bra{g}_2$, is assumed factorized from the beginning and has not yet interacted with the cavity. Thus, the procedure outlined in Fig.~\ref{fig: Methodology scheme} fully incorporates the measurement back-action effects on the cavity state, ensuring their complete consideration. This approach is also computationally more efficient than tracking all qubit states simultaneously, as it involves only a single qubit per interaction.

\subsection{Dealing with correlations}
\label{subsec: Dealing with correlations}
As highlighted in Eq.~(\ref{eq: Collapsed state |e> after measurement of 1st qubit}), the second FCS measurement on the $j$-th qubit completely destroys any qubit-cavity correlations that may have emerged. Thus, caution is warranted when applying the FCS if correlations significantly influence the process under study. However, in this context, FCS can be safely employed: we justify this as follows.

First, within the Jaynes-Cummings regime, qubit-cavity correlations arising from the charging process are negligible when using a Fock state cavity. This stems from the ideal dynamics occurring in the resonance regime, where properly tuned interaction times yield factorized qubit and cavity states. Second, this lack of significant correlations persists even under non-ideal conditions, such as slight detuning or minor timing errors, quantified explicitly in Sect.~\ref{app: Timing jitter} (see Fig.~\ref{fig: CORRELATIONS}).

By contrast, a two-point measurement scheme could induce non-negligible information loss in terms of erased correlations if other cavity states (e.g., Gaussian states) are used or substantial non-idealities (e.g., large timing errors) arise. In such cases, correlations may assume greater relevance (as illustrated in Fig.~\ref{fig: CORRELATIONS}), and omitting them could prevent the access to potentially useful resources. Nonetheless, in our scenario, we are interested solely in the energy extractable from a qubit measurement along the $z$-axis; thus, any energy flowed into correlations should be treated as lost. A detailed analysis in Sect.~\ref{subsec: Actual battery charging} further reveals that, even when measurements are performed onto the last charged qubit only, correlations arising during the charging of preceding qubits have no significant impact on charging performance. The FCS framework therefore accurately reflects the model's phenomenology, precisely describing outcomes from a realistic experimental quantum battery charging procedure.

\subsection{Operating-regime battery charging}
\label{subsec: Actual battery charging}
As previously noted, the methodology employed in our analysis can be interpreted as a battery characterization involving $2M$ measurements, two for each of the $M$ qubits.

To develop a strategy more representative of an operating regime, i.e., after the protocol has been fully characterized and no intermediate measurements are required, we now consider an alternative method for charging the stack of $M$ qubits. Specifically, we assume no measurements are performed on the first $M-1$ qubits, and only two measurements are taken on the $M$-th qubit. In this case, the state of the cavity and the $M-1$ qubits evolves undisturbed by projective measurements, potentially retaining correlations. We model this scenario without tracing over the qubits' degrees of freedom; instead, we numerically evolve the total joint state as a whole. Furthermore, the first FCS measurement on the $M$-th qubit has no effect on the overall state, since the $M$-th qubit's reduced state at the start of its cavity interaction is the factorized ground state.

\begin{figure}[htp!]
    \centering
\includegraphics[width=0.95\linewidth]{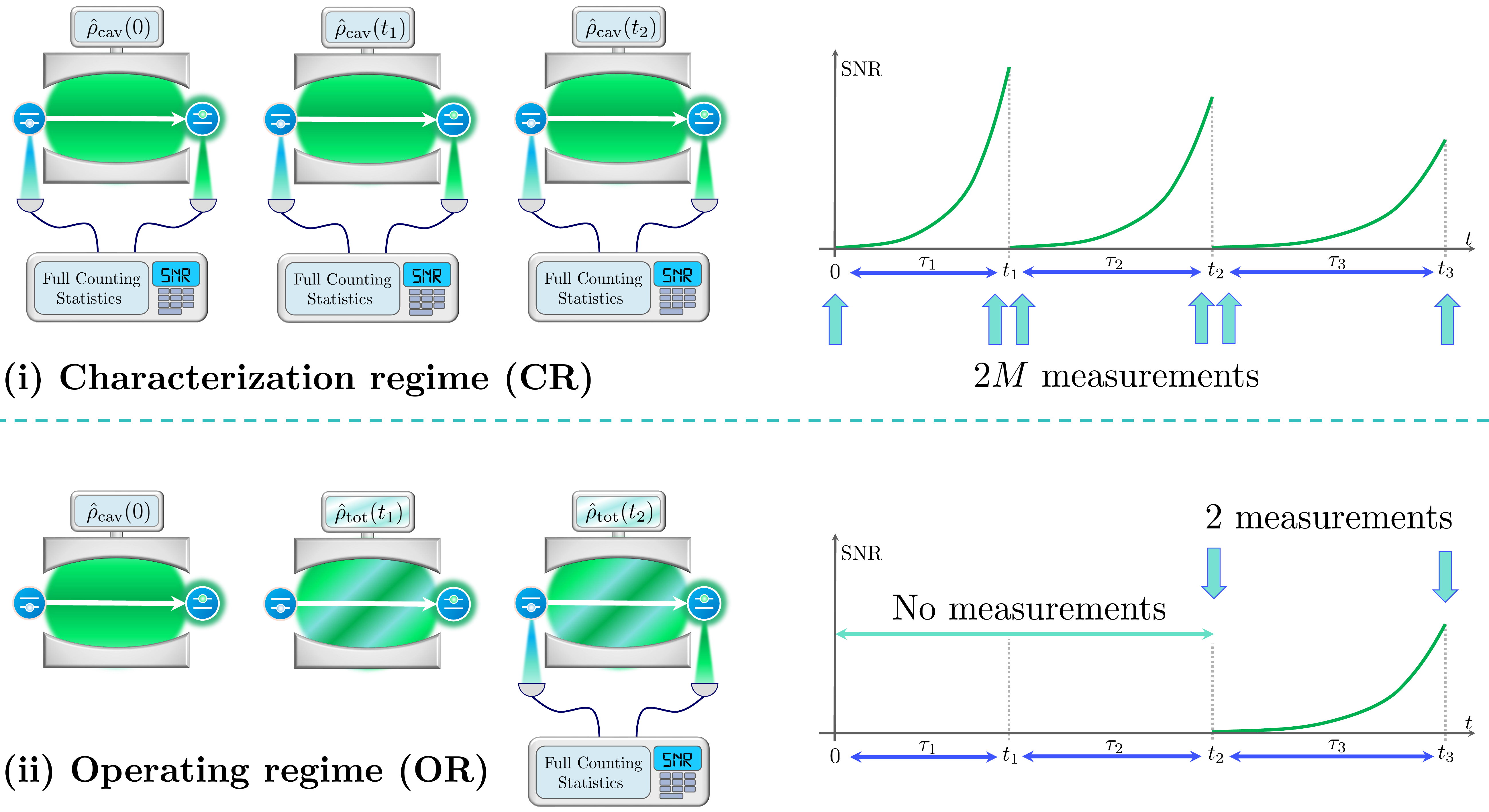}
    \caption{Comparison of (i) the characterization strategy used in our analysis with (ii) the operating-regime charging process, where no intermediate measurement is performed. The characterization involves $2M$ measurements, one per qubit, each perturbing the cavity state through back-action, modeled by tracing over the qubit's degree of freedom. In the operating regime, by contrast, no measurements precede the final qubit's charging: the total state of the qubits and cavity, $\hat{\rho}_{\text{tot}}(t)$, is generally non-separable. Only two measurements are then performed on the last qubit to assess its charging SNR via the FCS.}
    \label{fig: characterization vs operating regime}
\end{figure}

\begin{figure}[htp!]
    \centering
\includegraphics[width=0.55\linewidth]{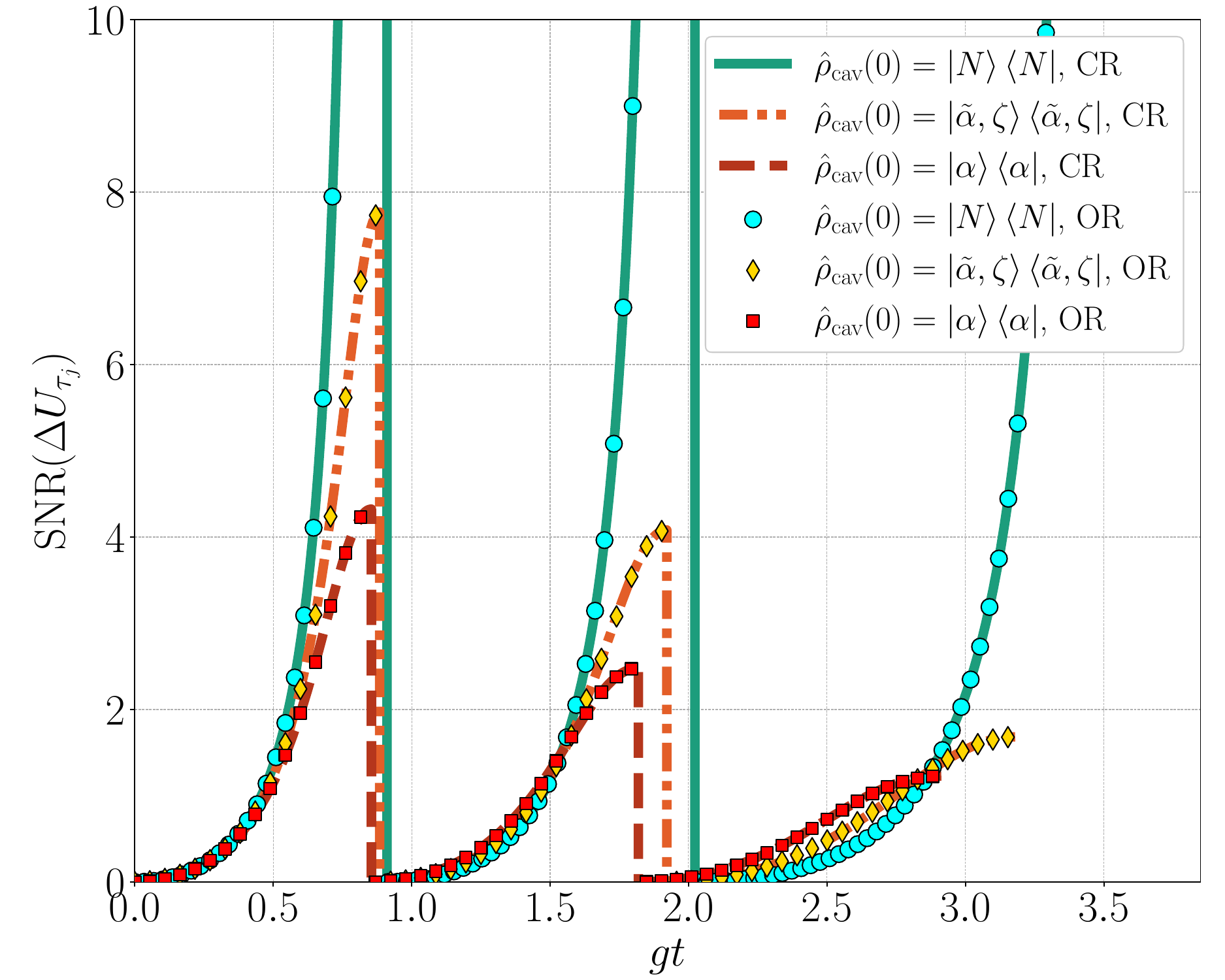}
    \caption{SNR computed in the characterization regime (CR) and in the operating regime (OR) for $M=3$ qubits. For brevity, each OR curve corresponds to three distinct protocols, one per peak. Specifically, the second peak describes the charging of two qubits with only the second one measured, while the third peak corresponds to a three-qubit charging process with only the third one measured. We note that the first OR peak is exactly equivalent to its CR counterpart, as they both involve charging and measuring the first qubit exclusively. As evident, no appreciable differences arise between CR and OR, indicating that any emergent correlations do not affect the battery charging performance. For the simulations, we used $N=\langle n \rangle = 3$, $\alpha = \sqrt{N}$, $\zeta = 0.6$, $\tilde{\alpha}\simeq 3.905$, $g/\omega_{\text{qub}}=10^{-2}$, and $\Delta\omega/\omega_{\text{qub}}=10^{-3}$.}
    \label{fig: comparison between characterization and operating-regime SNR}
\end{figure}

The differences between (i) the characterization methodology presented above and (ii) the battery charging in the operating regime are best visualized in Fig.~\ref{fig: characterization vs operating regime}. The quantitative comparison of the SNR obtained from the two strategies is reported in Fig.~\ref{fig: comparison between characterization and operating-regime SNR}, where no significant difference emerges between the methods. We therefore conclude that any spurious correlations arising from the absence of intermediate measurements during evolution do not impact the charging performance.

\clearpage
\section{Analytical solution of the two-qubit Tavis-Cummings model}
\label{app: Analytical solution of the two-qubit Tavis-Cummings model}
The evolution operator $\hat{\mathcal{U}}_{\text{TC}}(\tau)$ associated with the $M=2$ Tavis-Cummings Hamiltonian $\hat{H}_{\text{TC}}$, obtained by setting $A_1(t) = A_2(t) = 1$ $\forall t$ in Eq.~(1) in the main text, can be written explicitly as a $4\times 4$ matrix of cavity operators. A detailed derivation can be found in \cite{fujii2004TWOatoms, lu2025singleMISSINGTERM} (see in particular \cite{lu2025singleMISSINGTERM} for the terms $a_{24} = a_{34}$ in the exponentiation of the interaction Hamiltonian) for the resonant case, i.e., $\omega_{\text{qub}} = \omega_{\text{cav}} = \omega_0$. Here, we summarize the main results. \\

\par\noindent
The evolution operator from time $t=0$ to time $t=\tau$ can in general be written as:
\begin{equation}
    \label{eq: TC evolution operator, general}
\hat{\mathcal{U}}_{\text{TC}}(\tau) = e^{-\frac{i}{\hbar}\hat{H}_{\text{TC}} \tau} = e^{-i \frac{\omega_{0}}{2}\hat{\sigma}_z^{(1)} \tau  -i \frac{\omega_{0}}{2}\hat{\sigma}_z^{(2)} \tau -i\omega_{0} \hat{a}^\dagger\hat{a} \tau} e^{ -\frac{i}{\hbar}\hat{H}_{\text{int}} \tau}
\end{equation}
The tricky part of the computation consists of finding $e^{ -\frac{i}{\hbar}\hat{H}_{\text{int}} \tau}$, which can be expressed as \cite{fujii2004TWOatoms}:
\begin{equation}
    \label{eq: Exp of TC interaction H}
   e^{ -\frac{i}{\hbar}\hat{H}_{\text{int}} \tau} = \begin{pmatrix}
                a_{11} & a_{12} & a_{13} & a_{14} \\
                a_{21} & a_{22} & a_{23} & a_{24} \\
                a_{31} & a_{32} & a_{33} & a_{34} \\
                a_{41} & a_{42} & a_{43} & a_{44} \\
            \end{pmatrix} 
\end{equation}
where the matrix elements $a_{ij}$ are cavity operators defined by:

\begin{equation}
    \label{eq: TC Hint matrix elements definition, part 1}
\begin{split}
    a_{11} &= \frac{\hat{N}+2+(\hat{N}+1)\cos(g\tau\sqrt{2(2\hat{N} + 3)})}{2\hat{N}+3} \hspace{2.cm}
    a_{12} = a_{13} = -i\frac{\sin(g\tau\sqrt{2(2\hat{N} + 3)})}{\sqrt{2(2\hat{N} + 3)}} \hat{a} \\
    a_{14} &= \frac{-1 + \cos(g\tau\sqrt{2(2\hat{N} + 3)})}{2\hat{N}+3} \hat{a}^2 \hspace{3.4cm}
    a_{21} = a_{31} = -i\frac{\sin(g \tau\sqrt{2(2\hat{N} + 1)})}{\sqrt{2(2\hat{N} + 1)}} \hat{a}^\dagger \\
    a_{22} &= a_{33} = \frac{+1 + \cos(g\tau\sqrt{2(2\hat{N} + 1)})}{2}  \hspace{2.8cm}  a_{23} = a_{32} = \frac{-1 + \cos(g\tau\sqrt{2(2\hat{N} + 1)})}{2}  \\
    a_{24} &= a_{34} = -i\frac{\sin(g \tau\sqrt{2(2\hat{N} + 1)})}{\sqrt{2(2\hat{N} + 1)}} \hat{a}  \hspace{3.15cm}
    a_{41} = \frac{-1 + \cos(g\tau\sqrt{2(2\hat{N} - 1)})}{2\hat{N}-1} (\hat{a}^\dagger)^2  \\
   a_{42} &= a_{43} = -i\frac{\sin(g\tau\sqrt{2(2\hat{N} - 1)})}{\sqrt{2(2\hat{N} - 1)}} \hat{a}^\dagger  \hspace{3.cm}
    a_{44} = \frac{\hat{N}-1+\hat{N}\cos(g\tau\sqrt{2(2\hat{N} - 1)})}{2\hat{N}-1}
\end{split}
\end{equation}
and $\hat{N} = \hat{a}^\dagger\hat{a}$ represents the cavity number operator. Thus, by defining $\hat{\sigma}_z =$ \begin{scriptsize}$ {\begin{pmatrix}1 & 0 \\ 0 & -1
 \end{pmatrix}}$\end{scriptsize}, $\hat{\sigma}_z^{(1)} = \hat{\sigma}_z \otimes \hat{\mathds{I}}_2  \otimes \hat{\mathds{I}}_{\text{cav}}$, and $\hat{\sigma}_z^{(2)} = \hat{\mathds{I}}_1 \otimes \hat{\sigma}_z \otimes \hat{\mathds{I}}_{\text{cav}} $ (being $\hat{\mathds{I}}_i$ and $\hat{\mathds{I}}_{\text{cav}}$ the identity operators on the $i$-th qubit Hilbert space and on the cavity Hilbert space, respectively), one can write:
\begin{equation}
    \label{TC evolution operator, explicit}
       \hat{\mathcal{U}}_{\text{TC}}(\tau) = e^{-i\omega_{0}\hat{N} \tau} \begin{pmatrix}
                a_{11}e^{-i \omega_{0} \tau} & a_{12}e^{-i \omega_{0} \tau} & a_{13}e^{-i \omega_{0} \tau} & a_{14}e^{-i \omega_{0} \tau} \\
                a_{21} & a_{22} & a_{23} & a_{24} \\
                a_{31} & a_{32} & a_{33} & a_{34} \\
                a_{41}e^{+i \omega_{0} \tau} & a_{42}e^{+i \omega_{0} \tau} & a_{43}e^{+i \omega_{0} \tau} & a_{44}e^{+i \omega_{0} \tau} \\
            \end{pmatrix}     
\end{equation}

\section{Single-qubit and two-qubit energy statistics in the TC model}
\label{app: Single-qubit vs two-qubit energy statistics in the TC model}
Given the explicit form of $\hat{\mathcal{U}}_{\text{TC}}(\tau)$, we can exploit the FCS as in the preliminary work \cite{rinaldi2025reliable} to find an analytical expression for the statistics of the energy injected in a single qubit, as well as in both of them. To find the energy statistics of the single qubit (for simplicity, we focus on the first qubit, denoted as Q1), we have to compute the tilted operator:
\begin{equation}
    \label{eq: Tilted operator, single qubit energy}
    \hat{\mathcal{U}}^{\text{Q1}}_{\chi}(\tau) = e^{i\frac{\chi}{2} \frac{\hbar\omega_{0}}{2}\hat{\sigma}_z^{(1)}} \hat{\mathcal{U}}_{\text{TC}}(\tau) e^{-i\frac{\chi}{2} \frac{\hbar\omega_{0}}{2}\hat{\sigma}_z^{(1)}}
\end{equation}
\begin{equation}
    \label{TC evolution operator, tilted, single qubit, explicit}
       \hat{\mathcal{U}}^{\text{Q1}}_{\chi}(\tau) = e^{-i\omega_{0}\hat{N} \tau}   \begin{pmatrix}
                a_{11}e^{-i \omega_{0} \tau} & a_{12}e^{-i \omega_{0} \tau} & a_{13, \chi}e^{-i \omega_{0} \tau} & a_{14,\chi}e^{-i \omega_{0} \tau} \\
                a_{21} & a_{22} & a_{23, \chi} & a_{24,\chi} \\
                a_{31,-\chi} & a_{32,-\chi} & a_{33} & a_{34} \\
                a_{41,-\chi}e^{+i \omega_{0} \tau} & a_{42, -\chi}e^{+i \omega_{0} \tau} & a_{43}e^{+i \omega_{0} \tau} & a_{44}e^{+i \omega_{0} \tau} \\
            \end{pmatrix}     
\end{equation}
where $a_{ij,\pm\chi} = a_{ij} e^{\pm i\frac{\chi}{2}\hbar\omega_{0}}$. Analogously, the adjoint evolution operator modified with the parameter $-\chi$ is:
\begin{equation}
    \label{eq: U_-chi^dag for the Tavis-Cummings model, first qubit}            \hat{\mathcal{U}}^{\dagger\,\text{Q1}}_{-\chi}(\tau) =
         \begin{pmatrix}
                a^\dagger_{11}  e^{+i \omega_{0} \tau} &   a^\dagger_{21}  &  a^\dagger_{31,-\chi} &  a^\dagger_{41,-\chi} e^{-i \omega_{0} \tau}  \\
                  a^\dagger_{12} e^{+i \omega_{0} \tau} &  a^\dagger_{22} &  a^\dagger_{32, -\chi} &   a^\dagger_{42, -\chi} e^{-i \omega_{0} \tau} \\
                   a^\dagger_{13,\chi}e^{+i \omega_{0} \tau} &  a^\dagger_{23, \chi} &  a^\dagger_{33} & a^\dagger_{43} e^{-i \omega_{0} \tau}  \\
                    a^\dagger_{14,\chi} e^{+i \omega_{0} \tau} &   a^\dagger_{24,\chi} &  a^\dagger_{34} &  a^\dagger_{44} e^{-i \omega_{0} \tau} \\
                 \end{pmatrix}  e^{+i\omega_{0}\hat{N} \tau} 
\end{equation} 
being again $a^\dagger_{ij,\pm\chi} = a^\dagger_{ij} e^{\pm i\frac{\chi}{2}\hbar\omega_{0}}$. Now, if we assume that the qubits are initiallt prepared in the collective ground state $\ket{g}_1\bra{g}_1 \otimes \ket{g}_2\bra{g}_2 \equiv \ket{gg}\bra{gg}$ and that the initial cavity state is $\hat{\rho}_{\text{cav}}(0)$, we can compute the generating function associated with the statistics of the energy exchanged by the \textit{first qubit only} as follows:
\begin{equation}
    \label{eq: Generating function for TC model, first qubit}
    \begin{split}
            \mathcal{G}_{1}(\chi,\tau) =& \text{ Tr} \bigl\{ \hat{\mathcal{U}}^{\text{Q1}}_{\chi}(\tau) \bigl( \ket{gg}\bra{gg}  \otimes \hat{\rho}_{\text{cav}}(0) \bigr) \hat{\mathcal{U}}^{\dagger\,\text{Q1}}_{-\chi}(\tau) \bigr\} \\
            =& \text{ Tr}\{ a_{14} \hat{\rho}_{\text{cav}}(0)  a^\dagger_{14}e^{i\chi\hbar\omega_0} 
            + a_{24} \hat{\rho}_{\text{cav}}(0)  a^\dagger_{24}e^{i\chi\hbar\omega_0} 
            + a_{34} \hat{\rho}_{\text{cav}}(0)  a^\dagger_{34} + a_{44} \hat{\rho}_{\text{cav}}(0)  a^\dagger_{44} \}
    \end{split}
\end{equation}
By differentiating it with respect of $\chi$, we easily get the mean injected energy:
\begin{equation}
    \label{eq: Average injected energy in qubit 1}
    \begin{split}
    \langle \Delta U^{\text{Q1}}_\tau \rangle &= (-i)\frac{\partial}{\partial\chi}\mathcal{G}_{1}(\chi)\biggr\rvert_{\chi = 0} = \hbar\omega_0 \text{Tr}\{ a_{14} \hat{\rho}_{\text{cav}}(0)  a^\dagger_{14} +  a_{24} \hat{\rho}_{\text{cav}}(0)  a^\dagger_{24}\} 
    \end{split}
\end{equation}
and its variance: 
\begin{equation}
    \label{eq: Variance of injected energy in qubit 1}
    \begin{split}
    \text{var}(\Delta U^{\text{Q1}}_\tau) =&\,\, (-i)^2\frac{\partial^2}{\partial\chi^2}\mathcal{G}_{1}(\chi)\biggr\rvert_{\chi = 0} - \langle \Delta U^{\text{Q1}}_\tau \rangle^2 \\
    =&\,\, (\hbar\omega_0)^2 \text{Tr}\{ a_{14} \hat{\rho}_{\text{cav}}(0)  a^\dagger_{14} +  a_{24} \hat{\rho}_{\text{cav}}(0)  a^\dagger_{24}\}  -\langle \Delta U^{\text{Q1}}_\tau \rangle^2
    \end{split}
\end{equation}
If we use $\hat{\rho}_{\text{qub}}(0) = \ket{gg}\bra{gg}$ and $\hat{\rho}_{\text{cav}}(0) = \ket{N}\bra{N}$ as initial conditions, we obtain:
\begin{equation}
    \label{eq: Energy statistics of qubit 1}
    \begin{split}
\langle \Delta U^{\text{Q1}}_\tau \rangle &=\hbar\omega_0 \bigl[f(\tau,N) + h(\tau, N)\bigr] \\
\text{var}(\Delta U^{\text{Q1}}_\tau) &= (\hbar\omega_0)^2 \bigl[f(\tau,N) + h(\tau, N) \bigr] -\langle \Delta U^{\text{Q1}}_\tau \rangle^2
    \end{split}
\end{equation}
where the functions $f(\tau, N)$ and $h(\tau, N)$ are defined by:
\begin{equation}
\label{eq: Functions f and h, appendix}
\begin{split}
    f(\tau, N) &\equiv \text{Tr}\{ a_{14} \hat{\rho}_{\text{cav}}(0)  a^\dagger_{14}\} =\biggl[\frac{-1 + \cos(g\tau\sqrt{2(2N-1)})}{2N-1}\biggr]^2 N(N-1) \\
h(\tau, N) &\equiv \text{Tr}\{ a_{24} \hat{\rho}_{\text{cav}}(0)  a^\dagger_{24}\} = \frac{ \sin^2(g\tau\sqrt{2(2N-1)})}{2(2N-1)} N
\end{split}
\end{equation}
as in the main text.
The SNR is readily derived:
\begin{equation}
    \label{eq: SNR analytical for TC, appendix}
    \text{SNR}(\Delta U_\tau^{\text{Q1}}) = \frac{f(\tau,N) + h(\tau, N)}{1 - \bigl[ f(\tau,N) + h(\tau, N) \bigr]}
\end{equation}

\begin{figure}
    \centering
    \includegraphics[width=0.5\linewidth]{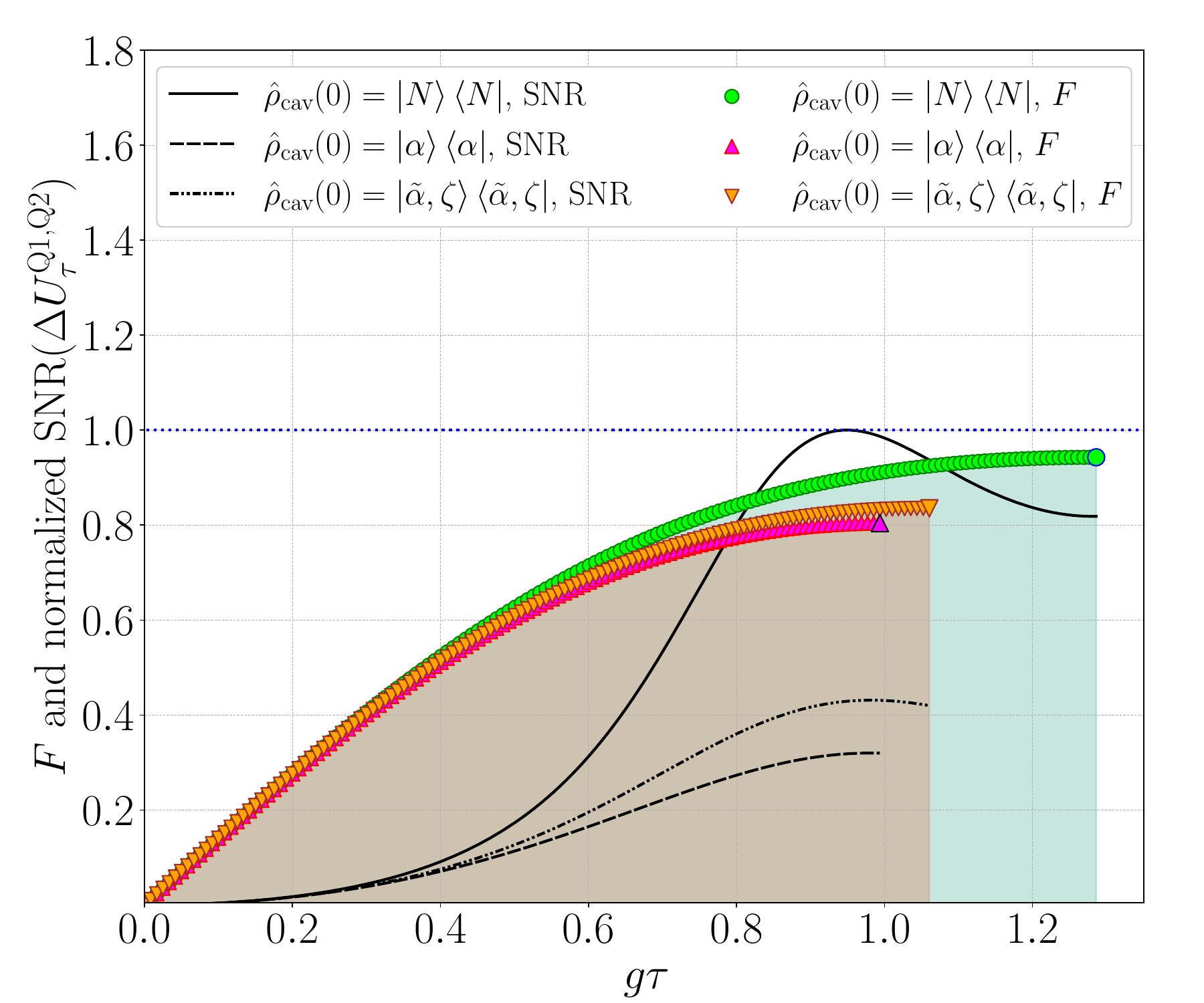}
    \caption{Fidelity $F$ (discrete points) calculated for the single-qubit state, compared to the SNR of the two-qubits energy exchange $\Delta U^{\text{Q1,Q2}}_\tau$, normalized to its maximum value (black lines). As it can be observed, the maximum of the SNR does not occur in corrispondence of the maximum of $F$: this means that the variance on the qubits collective energy is increased when the qubits are nearly charged. For that simulation, we used $\langle n \rangle = N=2$, $\alpha = \sqrt{2}$ for the coherent state, $\zeta = 0.6$ and $\tilde{\alpha} \simeq 2.3$ for the squeezed coherent state, and $g/\omega_{\text{qub}} = 10^{-2}$. Here, we considered the perfect-resonance condition, i.e., $\Delta\omega = 0$.}
    \label{fig: FIG10}
\end{figure}

Analogously, for what concerns the energy of the two qubits (labelled by Q1,Q2), we only have to compute the tilted operator by exponentiating the sum of the qubits free Hamiltonians:
\begin{equation}
    \label{eq: Tilted operator, two qubit energy}
    \hat{\mathcal{U}}^{\text{Q1,Q2}}_{\chi}(\tau) = e^{i\frac{\chi}{2} \frac{\hbar\omega_{0}}{2}(\hat{\sigma}_z^{(1)} + \hat{\sigma}_z^{(2)})} \hat{\mathcal{U}}_{\text{TC}}(\tau) e^{-i\frac{\chi}{2} \frac{\hbar\omega_{0}}{2}(\hat{\sigma}_z^{(1)} + \hat{\sigma}_z^{(2)})}
\end{equation}
This originates the following tilted operator:
\begin{equation}
    \label{TC evolution operator, tilted, two qubits, explicit}
    \begin{split}
       \hat{\mathcal{U}}^{\text{Q1,Q2}}_{\chi}(\tau) = e^{-i\omega_{0}\hat{N} \tau} 
       \begin{pmatrix}
                a_{11}e^{-i \omega_{0} \tau} & a_{12, \frac{\chi}{2}}e^{-i \omega_{0} \tau} & a_{13, \frac{\chi}{2}}e^{-i \omega_{0} \tau} & a_{14,\chi}e^{-i \omega_{0} \tau} \\
                a_{21, -\frac{\chi}{2}} & a_{22} & a_{23} & a_{24, \frac{\chi}{2}} \\
                a_{31, -\frac{\chi}{2}} & a_{32} & a_{33} & a_{34, \frac{\chi}{2}} \\
                a_{41,-\chi}e^{+i \omega_{0} \tau} & a_{42, -\frac{\chi}{2}}e^{+i \omega_{0} \tau} & a_{43, -\frac{\chi}{2}}e^{+i \omega_{0} \tau} & a_{44}e^{+i \omega_{0} \tau}
            \end{pmatrix}     
    \end{split}
\end{equation}
as well as:
\begin{equation}
    \label{eq: U_-chi^dag for the Tavis-Cummings model, two qubits}
    \begin{split}
\hat{\mathcal{U}}^{\dagger\,\text{Q1,Q2}}_{-\chi}(\tau) =
         \begin{pmatrix}
                a^\dagger_{11}  e^{+i \omega_{0} \tau} &   a^\dagger_{21, -\frac{\chi}{2}}  &  a^\dagger_{31, -\frac{\chi}{2}} &  a^\dagger_{41,-\chi} e^{-i \omega_{0} \tau}  \\
                  a^\dagger_{12, \frac{\chi}{2}} e^{+i \omega_{0} \tau} &  a^\dagger_{22} &  a^\dagger_{32} &   a^\dagger_{42, -\frac{\chi}{2}} e^{-i \omega_{0} \tau} \\
                   a^\dagger_{13, \frac{\chi}{2}}e^{+i \omega_{0} \tau} &  a^\dagger_{23} &  a^\dagger_{33} & a^\dagger_{43, -\frac{\chi}{2}} e^{-i \omega_{0} \tau}  \\
                    a^\dagger_{14,\chi} e^{+i \omega_{0} \tau} &   a^\dagger_{24, \frac{\chi}{2}} &  a^\dagger_{34, \frac{\chi}{2}} &  a^\dagger_{44} e^{-i \omega_{0} \tau} \\
                 \end{pmatrix} e^{+i\omega_{0}\hat{N} \tau} 
    \end{split}
\end{equation} 
where $a_{ij,\pm\frac{\chi}{2}} = a_{ij} e^{\pm i\frac{\chi}{2}\hbar\omega_{0}}$ and $a_{ij,\pm\chi} = a_{ij} e^{\pm i\chi\hbar\omega_{0}}$.
Again, by initializing the qubits in their ground state, the generating function related to the statistics of the energy exchanged by the \textit{two qubits as a whole} turns out to be:
\begin{equation}
    \label{eq: Generating function for TC model, two qubits}
    \begin{split}
            \mathcal{G}_{12}(\chi,\tau) =& \text{ Tr} \bigl\{ \hat{\mathcal{U}}^{\text{Q1,Q2}}_{\chi}(\tau) \bigl( \ket{gg}\bra{gg}  \otimes \hat{\rho}_{\text{cav}}(0) \bigr) \hat{\mathcal{U}}^{\dagger\,\text{Q1,Q2}}_{-\chi}(\tau) \bigr\} \\
            =& \text{ Tr}\{ a_{14} \hat{\rho}_{\text{cav}}(0)  a^\dagger_{14}e^{i2\chi\hbar\omega_0} + a_{24} \hat{\rho}_{\text{cav}}(0)  a^\dagger_{24}e^{i\chi\hbar\omega_0} + a_{34} \hat{\rho}_{\text{cav}}(0)  a^\dagger_{34}e^{i\chi\hbar\omega_0} + a_{44} \hat{\rho}_{\text{cav}}(0)  a^\dagger_{44} \}
    \end{split}
\end{equation}
As before, we straightforwardly find the energy statistics as:
\begin{equation}
    \label{eq: Average injected energy in two qubits}
    \begin{split}
    \langle \Delta U^{\text{Q1,Q2}}_\tau \rangle = (-i)\frac{\partial}{\partial\chi}\mathcal{G}_{12}(\chi)\biggr\rvert_{\chi = 0} = 2\hbar\omega_0 \text{Tr}\{ a_{14} \hat{\rho}_{\text{cav}}(0)  a^\dagger_{14}\} + \hbar\omega_0 \text{Tr}\{ a_{24} \hat{\rho}_{\text{cav}}(0)  a^\dagger_{24} + a_{34} \hat{\rho}_{\text{cav}}(0)  a^\dagger_{34} \}    
    \end{split}
\end{equation}
and: 
\begin{equation}
    \label{eq: Variance of injected energy in two qubits}
    \begin{split}
    \text{var}(\Delta U^{\text{Q1,Q2}}_\tau) &= (-i)^2\frac{\partial^2}{\partial\chi^2}\mathcal{G}_{12}(\chi)\biggr\rvert_{\chi = 0} - \langle \Delta U^{\text{Q1,Q2}}_\tau \rangle^2 \\
    &= 4(\hbar\omega_0)^2 \text{Tr}\{ a_{14} \hat{\rho}_{\text{cav}}(0)  a^\dagger_{14}\} + (\hbar\omega_0)^2 \text{Tr}\{ a_{24} \hat{\rho}_{\text{cav}}(0)  a^\dagger_{24} + a_{34} \hat{\rho}_{\text{cav}}(0)  a^\dagger_{34} \}  -\langle \Delta U^{\text{Q1,Q2}}_\tau \rangle^2
    \end{split}
\end{equation}
We note that, as expected, $\langle \Delta U^{\text{Q1,Q2}}_\tau \rangle = 2 \langle \Delta U^{\text{Q1}}_\tau \rangle$: indeed, since $a_{24}=a_{34}$, the last two terms in Eq.~(\ref{eq: Average injected energy in two qubits}) sum up to $2  \hbar\omega_0 \text{Tr}\{ a_{24} \hat{\rho}_{\text{cav}}(0)  a^\dagger_{24}\}$. On the other hand, the variances of the overall exchanged energy is not directly proportional to the one of the single qubit energy. This fact is reflected on the SNR of the two-qubit energy exchange: indeed, this reaches a peak before (and not in correspondence of) the SNR associated with the energy of the first qubit. This is evident from Fig.~\ref{fig: FIG10}, where the maximum of $\text{SNR}(\Delta U^{\text{Q1,Q2}}_\tau)$ does not coincide with the maximum of the single-qubit fidelity $F$.  \\

\par\noindent
To conclude this section, we note that the general $M$-qubit case can be approached numerically by solving Eq.~(8) 
in the main text. More generally, we can write the tilted Hamiltonian as $\hat{H}_{\text{TC}, \chi} = \hat{H}_{\text{free}} + \hat{H}^{\text{Q1}, ... \text{QM}}_{\text{int}, \chi}$, where $\hat{H}^{\text{Q1}, ... \text{QM}}_{\text{int}, \chi}$ is the $M$-qubit interaction Hamiltonian, tilted with the total free $M$-qubit Hamiltonian, i.e., 
\begin{equation}
    \label{eq: Tilted Hamiltonian for Tavis-Cummings, M qubits, part 1}
             \hat{H}^{\text{Q1}, ... \text{QM}}_{\text{int}, \chi} = e^{i\frac{\chi}{2}\hat{H}_{\text{free}}^{\text{qub}}} \hat{H}_{\text{int}} e^{-i\frac{\chi}{2}\hat{H}_{\text{free}}^{\text{qub}}} 
\end{equation}
with $\hat{H}_{\text{free}}^{\text{qub}} = \hbar\omega_{\text{qub}}\sum_{j=1}^M \frac{1}{2}\hat{\sigma}_z^{(j)}$. Since the qubits do not interact between each other, such operator turns out to be:
\begin{equation}
    \label{eq: Tilted Hamiltonian for Tavis-Cummings, M qubits, part 2}
    \begin{split}
             \hat{H}_{\text{int}, \chi}^{\text{Q1}, ... \text{QM}} &= \hbar g \sum_{j=1}^M (\hat{\sigma}_+^{(j)}\hat{a}e^{i\frac{\chi}{2}\hbar\omega_{\text{qub}}} + \hat{\sigma}_-^{(j)}\hat{a}^\dagger e^{-i\frac{\chi}{2}\hbar\omega_{\text{qub}}})
    \end{split}
\end{equation}
The numerical integration of Eq.~(8) 
with $\hat{H}_{\text{TC}, \chi} = \hat{H}_{\text{free}} + \hat{H}^{\text{Q1}, ... \text{QM}}_{\text{int}, \chi}$ then allows us to compute the statistics of the $M$-qubit energy exchange $\Delta U_\tau^{\text{Q1,...,QM}}$.

\section{Noise analysis}
\label{app: Additional noise analysis}
Here, we report the complete analysis of external factors that can diminish the Fock state advantage, ultimately corrupting the quantum features of the Fock state. Among them, we address two main factors of this kind: the issue of thermal noise, which we already anticipated in the main text, and the presence of state attenuation, which can be used to model imperfections in the cavity state preparation. Again, analogously to what we did to study the effects of thermal noise in the main text, we compare an attenuated Fock state of the field with a non-attenuated Gaussian state. To do so, we compute the same difference $D_{x, \langle n \rangle}$ of Eq.~(5) in the main text, where $x$ stands for a generic noise parameter (for instance, $x=\bar{n}_{\text{th}}$ for the thermal noise analysis). Moreover, we also account for possible limitations due to a suboptimal choice of the model parameters: specifically, we consider off-resonance conditions (i.e., when $\Delta\omega \neq 0$) and qubits which, instead of being prepared in their ground state, are initialized in low-temperature thermal states. In these two situations, both the Fock state and the Gaussian one are subject to the same choice of parameters. \\

\subsection{Thermal noise}
\label{subsec: Thermal noise}
The presence of noise in the initial cavity state can be a limiting factor for the protocol performances. As already presented in the main text, the first kind of noise that we considered is the one caused by the presence of thermal photons. To provide further details, we notice that the `thermalized' Fock state is modeled as a mixed state $\hat{\rho}_{\text{Fock}}^{\text{Th}} = \sum_{m=0}^{\infty} p_{m}(\bar{n}_{\text{th}}) \ket{m}\bra{m}$, where the probability distribution $p_m(\bar{n}_{\text{th}}, N)$ can be computed by exploiting a result derived in Ref.~\cite{ritboon2022sequential}, reading:
\begin{equation}
\label{eq: Fock state with thermal noise, prob distribution}
\begin{split}
         p_m(\bar{n}_{\text{th}}, N) &= \frac{m!}{N!\bar{n}_{\text{th}}} \sum_{k,k'=0}^m \frac{(-1)^{k+k'}}{k!k'!}\begin{pmatrix}
         N \\ m-k
     \end{pmatrix} \\
     &\times \begin{pmatrix}
         N \\ m-k'
     \end{pmatrix} (k+k'+N-m)! \\
     &\times \biggl(\frac{\bar{n}_{\text{th}}}{\bar{n}_{\text{th}}+1}\biggr)^{(k+k'+N-m+1)}
\end{split}
\end{equation}
The consequences of the presence of spurious thermal photons in the initial cavity state are reported in Fig.3 of main text, where they are also discussed.

\subsection{State attenuation}
\label{subsec: State attenuation}
\begin{figure}[ht]
    \centering
    \includegraphics[width=0.48\textwidth]{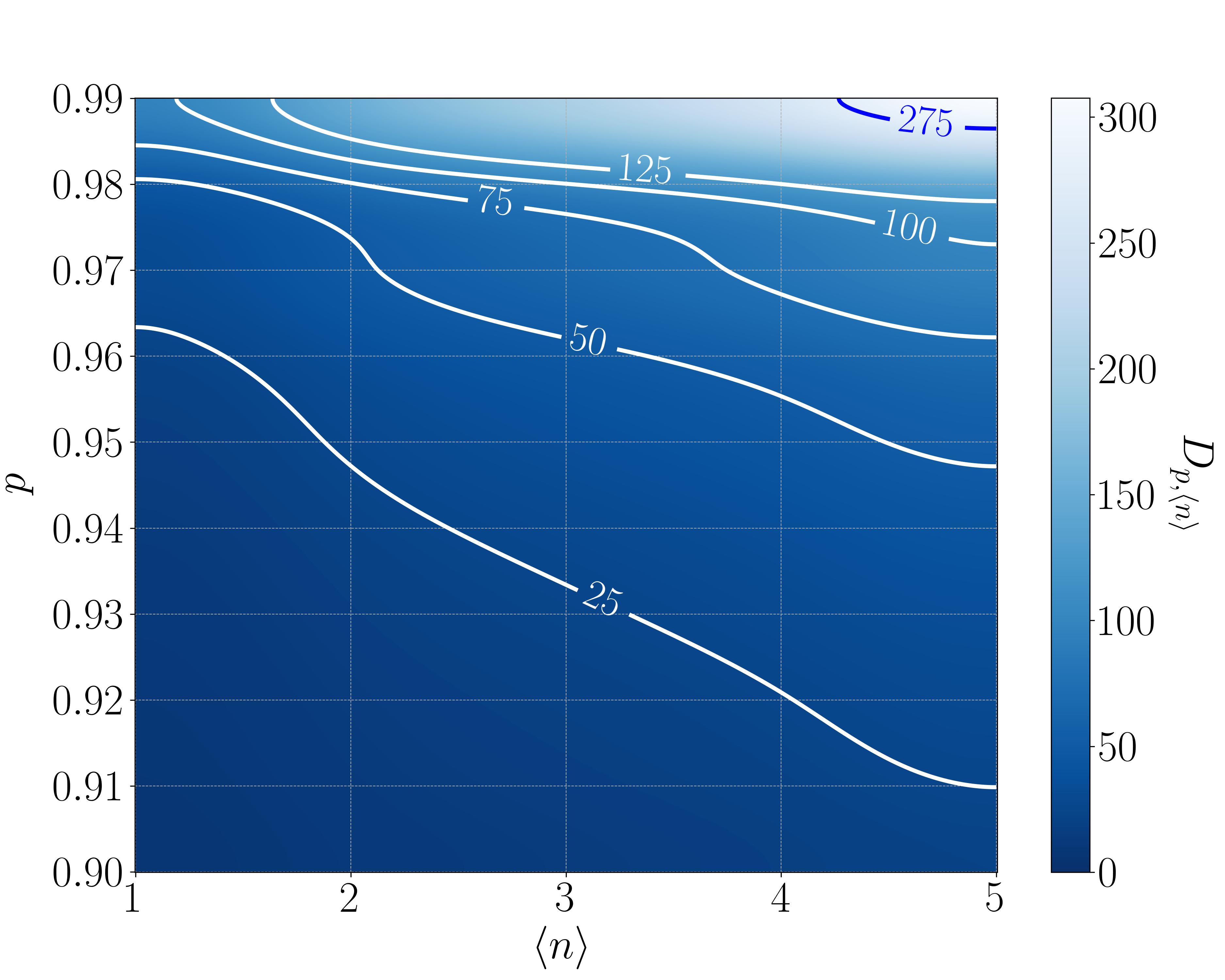}
    \caption{Effect of the cavity state attenuation, quantified by the probability $p$ of preparing the Fock state $\ket{N}\bra{N}$. If $p=1$, the state is not attenuated at all, while if $p=0$ the state is completely suppressed. The figure of merit is now $D_{p, \langle n \rangle}$, thus depending on $p$ and $\langle n \rangle$. Although the advantage quickly decreases as $p<1$, there is still an appreciable difference $D_{p, \langle n \rangle}\sim 25$ for $p\simeq 0.95$, which corresponds to the probability of preparing the Fock state $\ket{5}$ in an ion trap \cite{podhoraTHESIS} with current techniques. For that simulation, as for Fig.~3 in the main text, we used $N = \langle n \rangle \in\{1,...,5\}$ for the Fock state; $\zeta$ and $\tilde{\alpha}$ have been optimized to give the best performances for the squeezed coherent state. Besides, $g/\omega_{\text{qub}} = 10^{-2}$, and $\Delta\omega = 0$.
    }
    \label{fig: FIG8}
\end{figure}

Fock states are in general hard to be obtained: this implies that a realistic laboratory Fock state preparation can not be described by a pure state $\ket{N}\bra{N}$. Rather, the prepared state will be a mixture of $\ket{N}\bra{N}$ with probability $p$, plus other terms like $\ket{N-k}\bra{N-k}$, with $k$ integer. We model this sort of linear ``attenuation'' acting on the Fock state as if the state $\ket{N}\bra{N}$ experimented some losses, which can be described by a beam splitter \cite{leonhardt2003quantum}. The resulting state is the one transmitted through the beam splitter with probability $p$ (i.e., the square of the transmissivity $t$: $p = t^2$), and its mathematical expression is therefore:
\begin{equation}
    \label{eq: Transmitted state}
    \hat{\rho}_{\text{Fock}}^{\text{Attenuated}} = \sum_{k=0}^{N} \begin{pmatrix}
        N \\ k
    \end{pmatrix} p^k (1-p)^{N-k} \ket{k}\bra{k}
\end{equation}
with \begin{scriptsize}${\begin{pmatrix}
        N \\ k
    \end{pmatrix}} $\end{scriptsize} $ = \frac{N!}{k!(N-k)!}$. From Fig.~\ref{fig: FIG8} we can see that the difference $D_{p, \langle n \rangle}$ is heavily affected by the presence of attenuation; however, for $p\sim0.95$, the precision advantage is still considerable (for instance, see the case for $p=0.95$ and $\langle n \rangle=2$: $D_{p=0.95, \langle n \rangle=2} \simeq 25$). This is compatible with what has been experimentally observed in a trapped-ion device \cite{podhoraTHESIS}, where a Fock state $\ket{5}$ can be prepared with probability $p=0.95$. \\

\subsection{Suboptimal parameter choices}
\label{subsec: Suboptimal parameter choices}

\begin{figure}[htp]
    \centering
    \includegraphics[width=\textwidth]{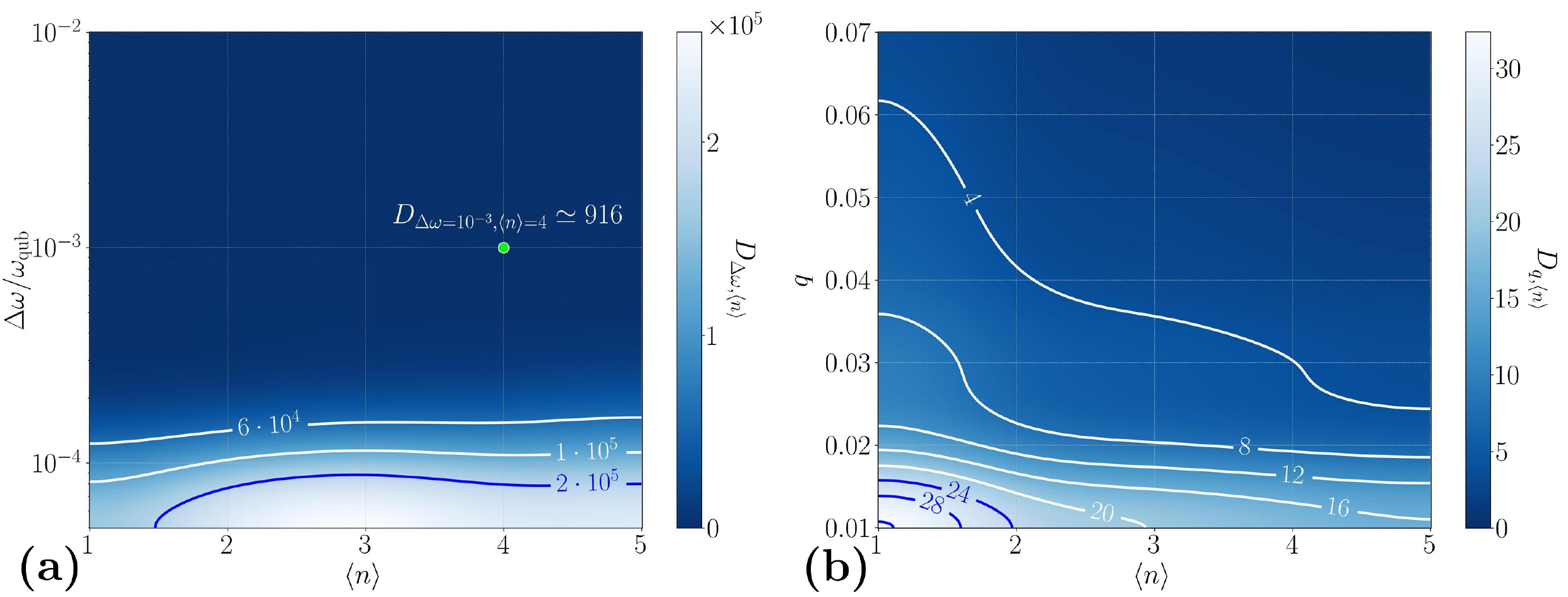}
    \caption{Effect of suboptimal choices of the model parameters, such as (a) the detuning $\Delta\omega = \omega_{\text{qub}} - \omega_{\text{cav}}$ and (b) the qubit populations $q$ such that $\hat{\rho}_{\text{qub}}(\tau=0) = q\ket{e}\bra{e} + (1-q)\ket{g}\bra{g}$. Note that, if $q=0$, the qubit initial state is exactly the ground state $\ket{g}\bra{g}$, while if $q = 0.5$, the qubit is in its maximally mixed state. The figures of merit are $D_{\Delta\omega, \langle n \rangle}$ (dependent on $\Delta\omega$ and $\langle n \rangle$) for case (a), and $D_{q, \langle n \rangle}$ (dependent on $q$ and $\langle n \rangle$) for case (b). The choice of the parameters is the same as in Fig.~\ref{fig: FIG8}. From (a), we can note that the Fock state advantage is very robust under the change of resonance conditions: indeed, non-negligible values of $\Delta\omega/\omega_{\text{qub}}$ (such as $\sim 10^{-3}$) allow for a conspicuous advantage. On the other hand, from (b) we see that the advantage is dramatically decreased in the case of non-zero qubit populations $q$. To maintain an appreciable advantage, thus, an experimental realization should assure the usage of low-temperature (nearly-ground state) qubits, for which at least $q \lesssim 0.01$. Besides, as the number of qubits increases, the advantage decreases in presence of a nonzero $q$, making this a considerable challenge to overcome when addressing the realization of a scalable quantum battery.
    }
    \label{fig: FIG9}
\end{figure}

We finally consider what happens to the charging performances if we vary the model parameters, thus missing the optimal conditions that allow for a perfect qubit charging. The first parameter that we change is the detuning $\Delta\omega = \omega_{\text{qub}} - \omega_{\text{cav}}$ between qubit and cavity. The second one consists in the population $q$ of the qubit excited level: we assume that the qubit is prepared in a low-temperature state defined as the mixture $q\ket{e}\bra{e} + (1-q)\ket{g}\bra{g}$. Fig.~\ref{fig: FIG9} reports the results for both the parameters. Clearly, the presence of non-perfect resonance is not a problem for the battery performances: indeed, from Fig.~\ref{fig: FIG9}a it is possible to see that the difference $D_{\Delta\omega, \langle n \rangle}$ is extremely high even for non-negligible detuning, such as $\Delta\omega/\omega_{\text{qub}}= 10^{-3}$. As an example, as reported in the figure, for $\Delta\omega/\omega_{\text{qub}}= 10^{-3}$ and $\langle n \rangle = 4$, we get for instance $D_{\Delta\omega, \langle n \rangle}\simeq 916$. The true limiting factor, instead, is represented by the qubit populations $q$. Indeed, if we observe Fig.~\ref{fig: FIG9}b, we note that $q\sim0.01$ is sufficient to suppress the advantage down to $D_{q, \langle n \rangle}\sim 30$ for $\langle n \rangle = 1$. Moreover, $D_{q, \langle n \rangle}$ is further decreased as $\langle n \rangle$ increases: this means that the presence of a populated excited state in the qubits also affects the charging of the overall stack of qubits.

\section{Charging speed in sequential and parallel protocols}
\label{app: Charge speed in sequential and parallel protocols}
Another notable question regards the charging speed of a certain protocol, which is of interest in the study of quantum batteries. As can be noted by looking at Fig.~2 
in the main text,
the sequential charging protocol needs a longer time to be completed, if compared to the parallel one. Indeed, as expected, the time needed to charge the $j$-th qubit of the sequential stack depends on the number of photons present in the cavity at the beginning of the charging process: in particular, for a $N_j$-photon Fock state cavity ready to charge the $j$-th qubit, the charging time $\tau_j$ depends on $N_j$ as $\tau_j \propto 1/\sqrt{N_j}$ \cite{andolina2018charger, rinaldi2025reliable}. This means that the sequential protocol is slower than the parallel one (which naturally charges the whole stack of qubits simultaneously), especially for an increasing number of qubits $M$. For that reason, one may argue that this fact should be taken into account to define a genuine quantum advantage.  
\\
In that perspective, we made an `unfair' comparison where a realistic (noisy) sequential protocol competes with an ideal (noiseless) parallel protocol to charge $M=N$ qubits with a $N$-photon Fock state. In particular, for the sequential protocol, we choose some non-ideal initial conditions: the qubits are in a low-temperature termal state with population $q = 10^{-3}$, and they are not in perfect resonance with the cavity: $\Delta\omega/\omega_{\text{qub}} = 10^{-3}$. Besides, only for the sequential protocol, we account for the presence of thermal photons (as an example, we choose $\bar{n}_{\text{th}} = 10^{-2}$), as well as of attenuation (we choose $p=0.98$) in the initial cavity state. \\

\begin{figure}[htp!]
    \centering
\includegraphics[width=0.5\linewidth]{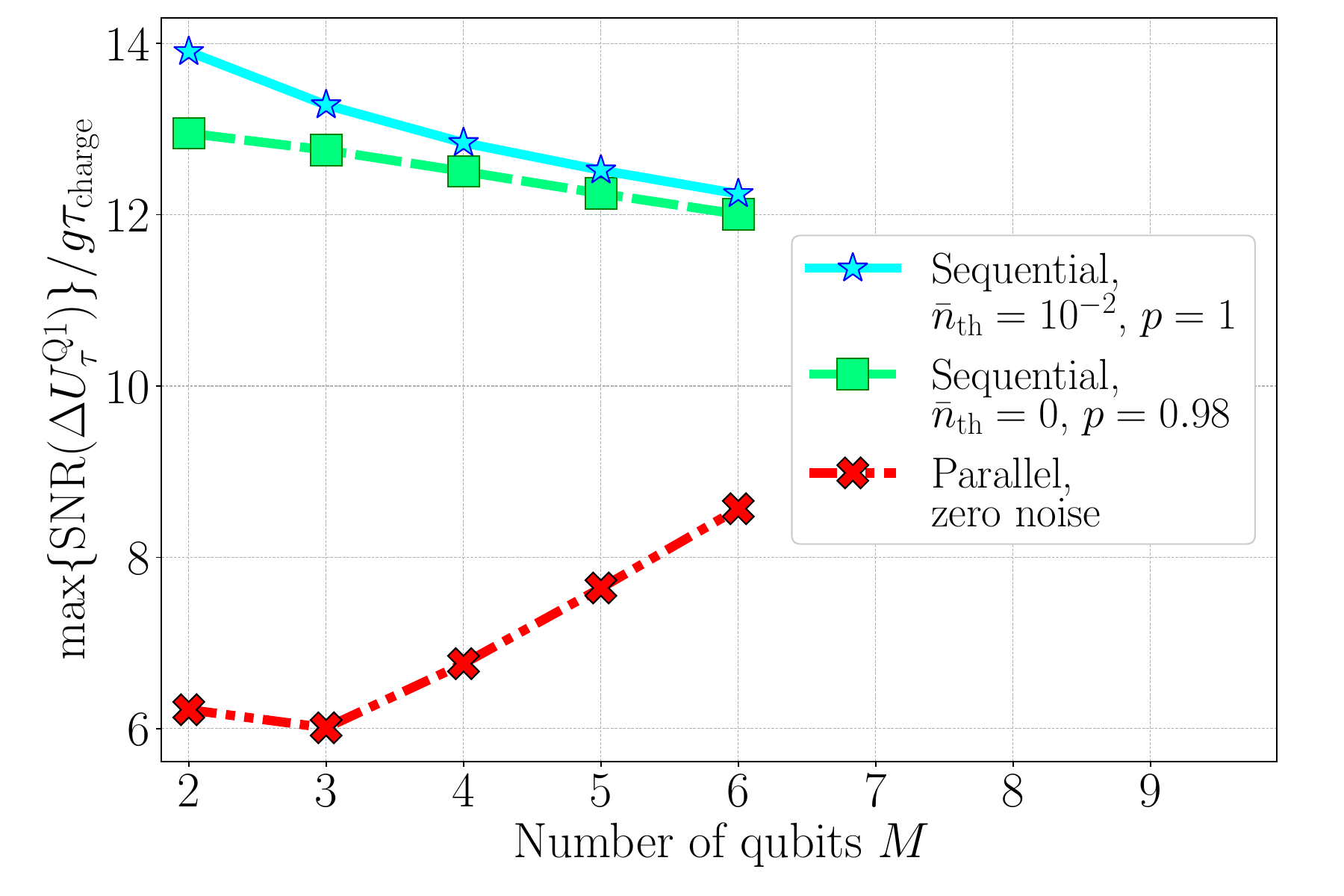}
    \caption{Worst-case comparison between realistic (noisy) sequential charging and ideal (noiseless) parallel charging of $M=N$ qubits with a $N$-photon Fock state cavity. While the parallel protocol (red crosses) is simulated under perfect conditions (i.e., the qubits are in their ground state at the beginning of the protocol, they are perfectly resonant with the cavity, and no noise is present), the sequential one is performed with suboptimal model parameters and in presence of noise. In particular, for the latter we choose $q = 10^{-3}$ and $\Delta\omega/\omega_{\text{qub}} = 10^{-3}$. The Figure displays the case with thermal noise but no attenuation ($\bar{n}_{\text{th}}=10^{-2}$ and $p=1$, blue stars) and the case with attenuation but no thermal noise ($\bar{n}_{\text{th}}=0$ and $p=0.98$, green squares). The points represent the average maximum SNR, divided by the time $\tau_{\text{charge}}$ required to perform the charging of $2 \leq M \leq 6$ qubits. The specific choice of parameters allows the realistic sequential protocol to achieve performances that are better than the parallel protocol. In spite of the fact that, for higher $M$, we expect the performances to be comparable to (if not worse than) the parallel charging, we point out that it is possible to arbitrarily increase the ones of the sequential charging by suppressing the noise and by choosing optimal model parameters.}
    \label{fig: FIG11}
\end{figure}

To account for the charging speed, we thus define $\tau_{\text{charge}}$ as the time needed to attain the maximum achievable charge in all the qubits. In the case of the parallel protocol, $\tau_{\text{charge}}$ is simply the time that corresponds to the maximum SNR obtained in the single-qubit charging: therefore we use such maximum SNR divided by the dimensionless quantity $g\tau_{\text{charge}}$ as a figure of merit. In the case of the sequential protocol, instead, $\tau_{\text{charge}}$ is the overall time needed to charge the entire chain of qubits, i.e., the time passed from the instant when the first qubit is coupled to the cavity, to the instant when the last qubit is decoupled from it. We average the SNR achieved for each single-qubit charging by summing all the SNR maxima, and then we divide by the number of qubits that have been charged. We finally divide such averaged SNR by $g\tau_{\text{charge}}$. \\
From Fig.~\ref{fig: FIG11} it is possible to see that the average SNR/time can be higher for the sequential protocol rather than the parallel one,  at least for a small number of qubits. However, in the limit of a high number of qubits $M$, the parallel protocol seems destined to overcome the sequential one in terms of average SNR/time. This fact is reasonable, because the parallel protocol allows for achieving more or less the same SNR for each $M$, but the process takes shorter and shorter charging times $\tau_{\text{charge}}$ as $M$ increases. On the other hand, the sequential protocol allows to reach a higher SNR on average, but $\tau_{\text{charge}}$ is longer as $M$ is increased. Nevertheless, the average SNR/time in the sequential protocol can be arbitrarily enhanced by optimizing the model parameters and by suppressing the noise. Furthermore, we presented here a worst-case comparison: the parallel protocol, indeed, was simulated under ideal conditions. Even in that case, however, the sequential protocol proved itself to be capable of overcoming the parallel one at least for a finite number of qubits. \\
Moreover, it shows an expected no-free-lunch behaviour: using better resources, we can charge more precisely all qubits only if we go slower. On the other hand, if we want to speed up charging with much less precision, lower resources are sufficient.

\section{Delving deeper into the sequential strategy: potential issues}
\label{app: Delving deeper into the sequential strategy: potential issues}
Here, we focus on the sequential charging protocol and analyze in details two possible factors which can limit the performance of the process: the presence of timing jitter when coupling the qubits to the cavity, and of a dissipation channel which destroys the coherences of the cavity state.

\subsection{Timing jitter errors: the effect of imprecise time tuning}
\label{app: Timing jitter}
Another possible source of errors, and therefore of a detrimental effect that can decrease the precision advantage in the sequential charging protocol, is the presence of a timing jitter phenomenon. In particular, one might wonder whether or not the precision advantage in the SNR is affected by an imprecise choice of the interaction time windows $\tau_j$. In this respect, we would like to contextualize the treatment within today's technological advancements. Over the last two decades, several scientific studies have focused on developing fast tunable couplings between integrated quantum systems. For instance, we refer to Ref.~\cite{chen2014qubit_FAST_TUNABLE_COUPLING}, where a fast on/off switch of coupling strengths on the order of nanoseconds has been demonstrated on a superconducting platform, where the dynamics usually occurs over a duration on the order of microseconds. Even more recently, it has been shown \cite{stehlik2021tunable} that nanosecond operations (again on superconducting devices) can be performed with at most $\sim6 \cdot 10^{-3}$ error per gate. With these results in mind, we investigated the sequential charging protocol by adding a timing jitter error to the charging times $\tau_j$. In practice, instead of simulating a perfect interaction time, we introduced an error $\epsilon$ such that each interaction time is a random number $\tau_j \pm \epsilon$. The error $\epsilon$ can occur according to a Gaussian probability distribution centered at 0 with a standard deviation $\sigma$. We quantified this standard deviation in terms of the optimal time needed to charge the first qubit with an $N$-photon Fock state, which reads \cite{rinaldi2025reliable}:
\begin{equation}
\tau_{\text{opt}} = \frac{\pi}{2\sqrt{g^2 N + \left(\frac{\Delta \omega}{2}\right)^2}}
\end{equation}
For $N=5$, and considering a coupling strength $g \sim 10$ MHz with a negligible detuning (achievable on superconducting platforms \cite{forn2019ultrastrong}), we get $\tau_{\text{opt}} \simeq \frac{\pi}{2\sqrt{5}} \cdot 10^{-7}$ s, which amounts to approximately 70 ns (comparable to the gate duration in \cite{stehlik2021tunable}). In practice, we accounted for $\sigma/\tau_{\text{opt}} \in \{10^{-3}, 10^{-2}, 10^{-1}, 5 \cdot 10^{-1}\}$. For each value of $\sigma/\tau_{\text{opt}}$ and for each cavity state among the Fock, coherent squeezed, and coherent states, we simulated the charging process 100 times. In each charging process, we accounted for the error $\epsilon$ on each of the five charging times $\tau_j$. After the $j$-th qubit charging process, we computed the average of the achieved SNR associated with the charging time $\tau_j$.

\begin{figure}[htp!]
    \centering
    \includegraphics[width=0.9\textwidth]{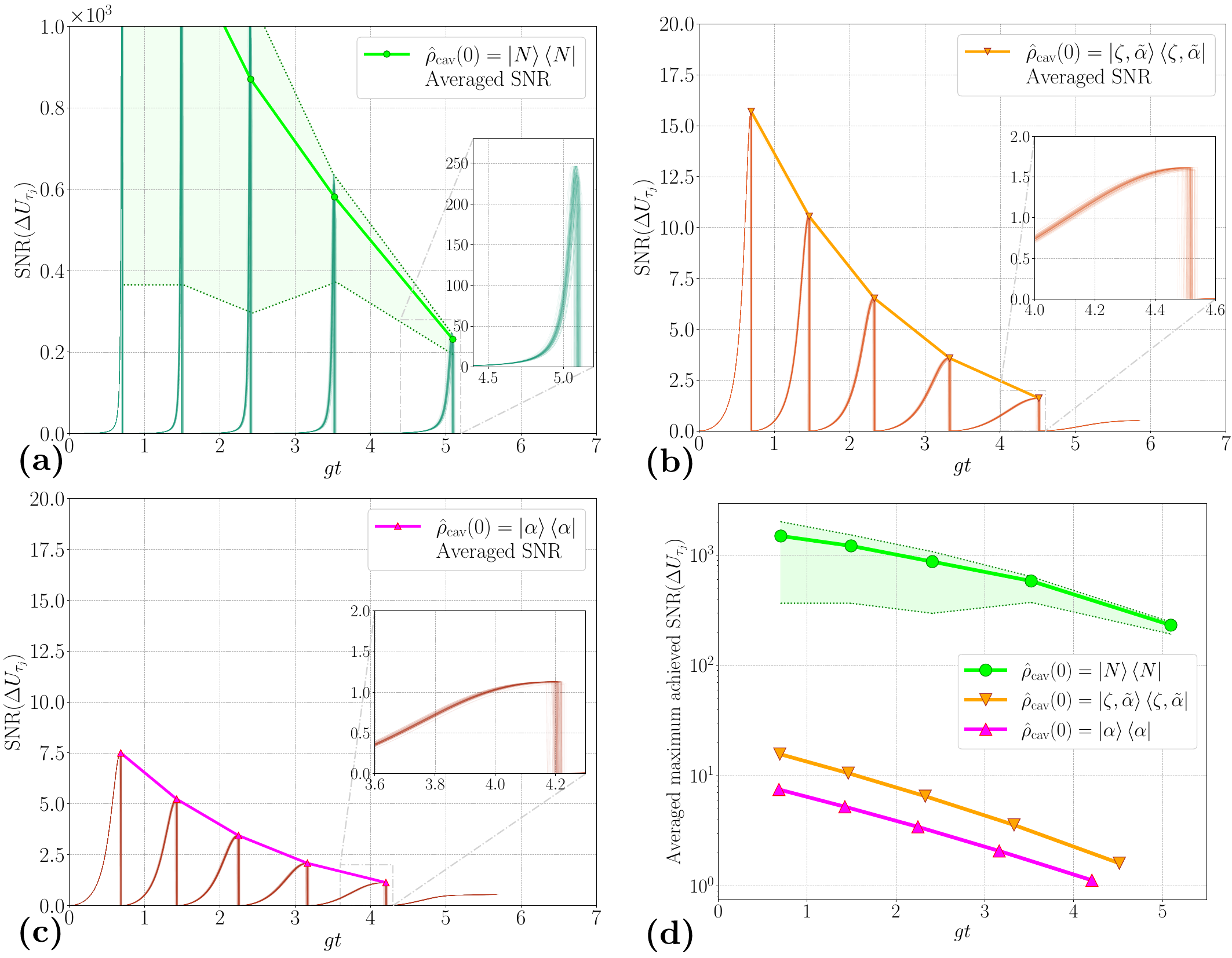}
    \caption{Accounting for timing jitter in the $M=5$ qubits sequential protocol. We simulated the charging process of 5 qubits with the possibility of making an error while tuning the charging time $\tau_j$. Each error $\epsilon$ such that the actual charging time is $\tau_j \pm \epsilon$ can randomly occur according to a Gaussian probability distribution centered in 0 and with standard deviation $\sigma$. We made 100 runs for each charging protocol, and we sampled the error $\epsilon$ for each charging time $\tau_j$ (that is, 5 times during each charging process). In this Figure, we explicitly show the results obtained with $\sigma/\tau_{\text{opt}} = 10^{-2}$, for the (a) Fock state, (b) coherent squeezed state, and (c) coherent state protocol. The insets magnify the SNR obtained for each of the 100 runs, to emphasize the fact that each line is originated from a stochastic choice of charging times. We then calculate the maximum SNR associated with the $j$-th qubit obtained for each run, and then we compute their average, represented by the markers and summarized in (d). Additionally, we also show the lowest and the highest SNR values obtained during the 100 runs for the Fock state protocol, to provide a general idea of the variability of the SNR itself (dashed lines and shaded region). Simulation parameters: $N = \langle n \rangle = 5$ for the Fock state; $\zeta = 0.6$ and $\tilde{\alpha} \simeq 2.14$ for the coherent squeezed state; $\alpha = \sqrt{N}$ for the coherent state, $g/\omega_{\text{qub}} = 10^{-2}$, and $\Delta\omega/\omega_{\text{qub}} = 10^{-3}$.} 
    \label{fig: TIME JITTER 1}
\end{figure}

\begin{figure}[htp!]
    \centering
    \includegraphics[width=0.55\textwidth]{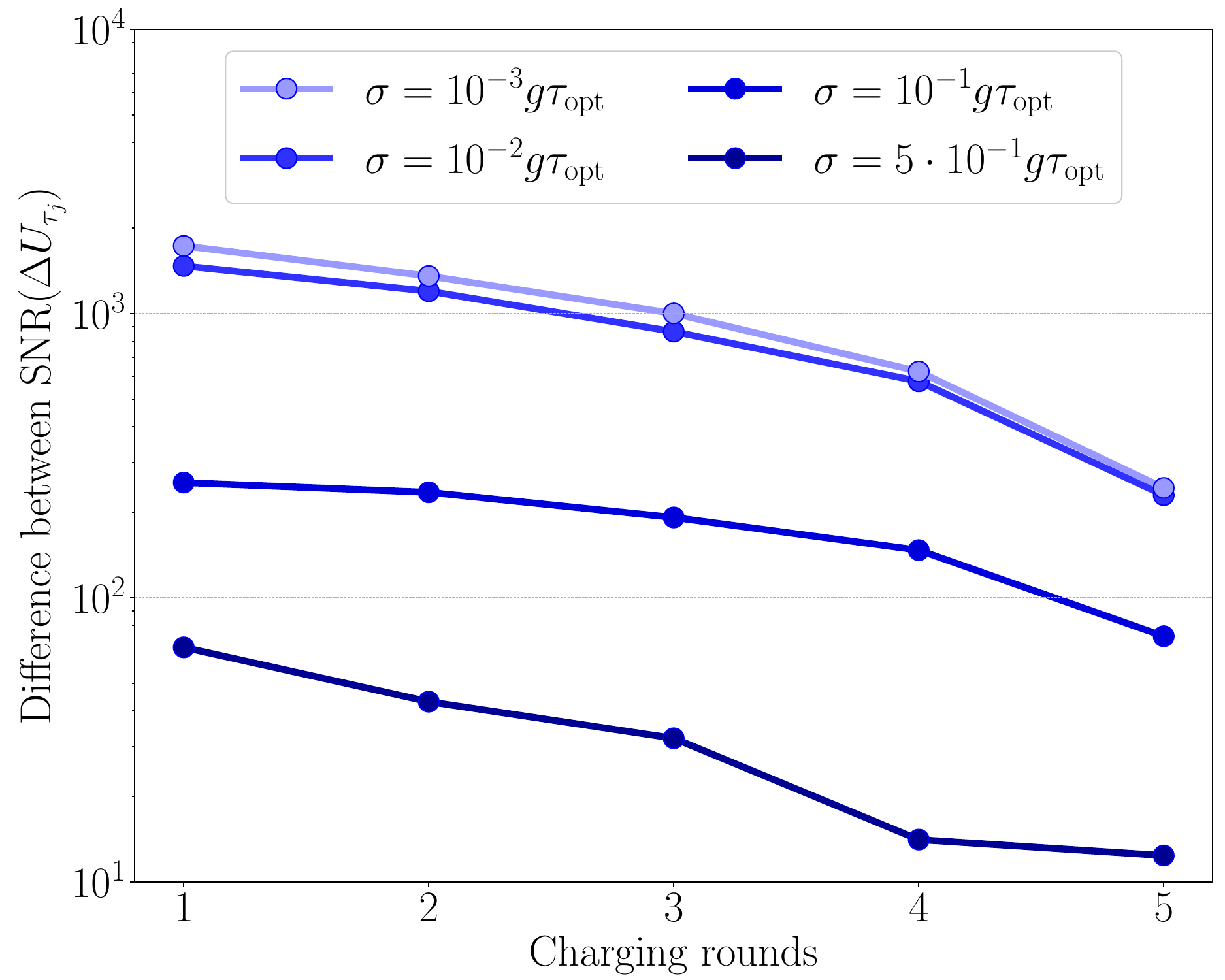}
    \caption{Influence of timing jitter on the $M=5$ qubits sequential protocol. Here we show the difference between the average maximum SNR obtained in Fig.~\ref{fig: TIME JITTER 1}d for the Fock state and for the coherent squeezed state, varying the width $\sigma$ of the Gaussian that governs the error probability. Standard deviations with magnitude $\sigma/\tau_{\text{opt}}\geq 10^{-1}$ are actually quite coarse, and indeed each of the 100 runs can return low or, in general, very variable SNR values. Nevertheless, on average the difference between the Fock state and the Gaussian state protocols is still positive and appreciable. The simulation parameters are the same as in Fig.~\ref{fig: TIME JITTER 1}.} 
    \label{fig: TIME JITTER 2}
\end{figure}

As observed from Fig.~\ref{fig: TIME JITTER 1} for $\sigma/\tau_{\text{opt}} = 10^{-2}$, and from Fig.~\ref{fig: TIME JITTER 2}, where the performances under different timing jitter magnitudes are compared, the Fock state protocol is still robust even under timing jitter errors, presenting remarkable resilience even for high error values. When the time-tuning precision becomes very coarse (such as for $\sigma/\tau_{\text{opt}} > 10^{-1}$), the performances of the Fock state can occasionally be comparable to those of the Gaussian state, but they still remain higher on average. We further point out that we carried out the analysis considering a slightly off-resonance condition ($\Delta \omega/\omega_{\text{qub}} = 10^{-3}$); therefore, the protocol is no longer relying on such an idealized assumption.

\subsection{Qubit-qubit correlations}
\label{app: Qubit-qubit correlations}
It is also important to account for the presence of emergent qubit-qubit correlations during the charging process. Their presence can indeed affect the charging performances and must therefore be properly investigated. On the other hand, understanding the role of correlations in the charging protocol can help shed light on the origin of the precision advantage provided by the use of a Fock state as a resource.

To begin with, let us assume perfect control over the tuning of the interaction times $\tau_j$. In the Fock-state cavity protocol, after each time window $\tau_j$, the reduced states of the $j$-th qubit and of the cavity are factorized due to the peculiar nature of the Jaynes-Cummings interaction. Therefore, a timely quantum measurement performed on the qubit will not affect the cavity state, and vice versa. In this idealized situation, no correlations can arise between the qubits: indeed, they never directly interact with each other and, furthermore, they do not interact via the mediation of the cavity, since it is uncorrelated with the previous qubit. This is reflected in the variance of the exchanged energy, which approaches zero as the interaction time tends to $\tau_j$.

However, one could argue that this argument no longer holds if the interaction time differs from $\tau_j$. In such a case, at the end of the charging process, the qubit and the cavity could be entangled, leading to a nonzero variance even in the Fock state scenario. We addressed this issue by focusing on a two-qubit sequential protocol, for which we monitored the overall state of the cavity and the two qubits during the entire charging process. To quantify the amount of correlations emerging between the two qubits during the charging process, we compute the quantum mutual information associated with the two-qubit joint state $\hat{\rho}_{\text{qub}}^{(1,2)}$:
\begin{equation}
\mathcal{I}(\hat{\rho}_{\text{qub}}^{(1,2)}) = S(\hat{\rho}_{\text{qub}}^{(1,2)}||\hat{\rho}_{\text{qub}}^{(1)}\otimes\hat{\rho}_{\text{qub}}^{(2)}),
\end{equation}
where $S(\hat{\rho}||\hat{\sigma}) = \text{Tr}\{\hat{\rho}\log\hat{\rho} - \hat{\rho}\log\hat{\sigma}\}$ is the quantum relative entropy between two states $\hat{\rho}$ and $\hat{\sigma}$, and $\hat{\rho}_{\text{qub}}^{(j)}$ is the reduced state of the $j$-th qubit. The quantity $\mathcal{I}(\hat{\rho}_{\text{qub}}^{(1,2)})$ accounts for the total amount of correlations \cite{plenio2005mutual_info}, both classical and quantum: it therefore represents a suitable quantifier of how much the two-qubit state cannot be factorized into the product of the two reduced states.

\begin{figure}[htp!]
    \centering
    \includegraphics[width=0.9\textwidth]{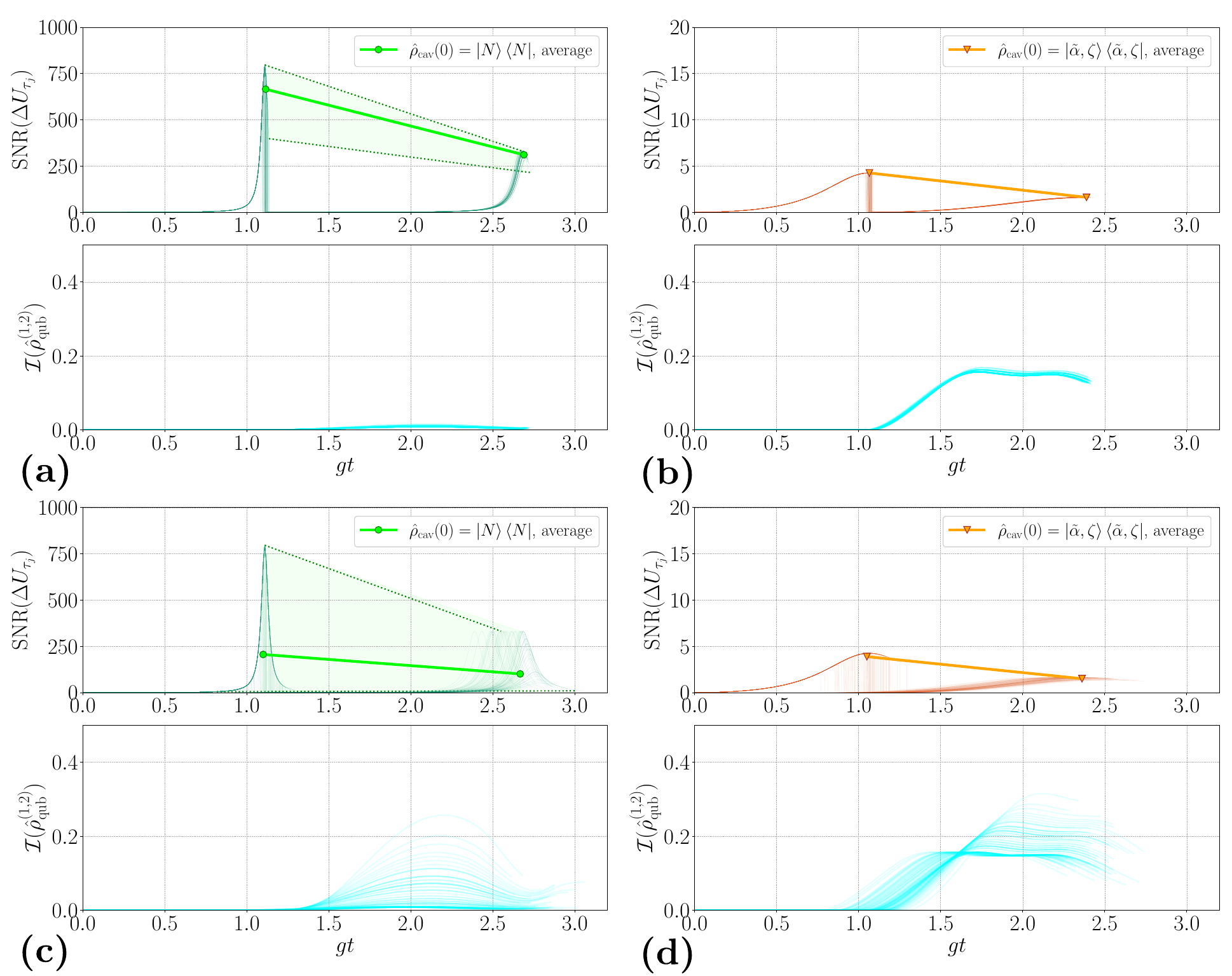}
    \caption{Emergence of qubit-qubit correlations in the $M=2$ qubits sequential protocol subject to timing jitter. We discuss the case for $\sigma/\tau_{\text{opt}} = 10^{-2}$ for (a) a Fock state and for (b) a squeezed coherent state protocol, as well as the case with a large timing jitter error, $\sigma/\tau_{\text{opt}} = 10^{-1}$, again for (c) a Fock state and (d) squeezed coherent state protocol. Each of these four plots is divided in two parts: the upper one shows the SNR obtained for 100 charging process simulations with the corresponding timing jitter error; the lower one presents the mutual information $\mathcal{I}(\hat{\rho}_{\text{qub}}^{(1,2)})$ associated with the two-qubit joint state, $\hat{\rho}_{\text{qub}}^{(1,2)}$, calculated for each of the 100 simulations. The mutual information quantifies the total amount of correlations (both classical and quantum) that arise between the two qubits. As we can note, in presence of reasonable timing jitter error, such as $\sigma/\tau_{\text{opt}} = 10^{-2}$, the presence of correlations due to the Fock state protocol is lower, if compared to the one associated with the squeezed coherent state, which is non-negligible. If the time tuning error is greater, for instance $\sigma/\tau_{\text{opt}} = 10^{-1}$, then spurious correlations appear also in the Fock state scenario, although with a high range of variability. The simulation procedure and parameters are the same as in Fig.~\ref{fig: TIME JITTER 1}.} 
    \label{fig: CORRELATIONS}
\end{figure}

As we did for the timing jitter analysis, we performed 100 simulations of the two-qubit charging process, employing both a Fock state and a squeezed coherent state cavity. We accounted for two possible timing jitter errors: $\sigma/\tau_{\text{opt}} = 10^{-2}$ and $\sigma/\tau_{\text{opt}} = 10^{-1}$. For each simulation, we computed the mutual information $\mathcal{I}(\hat{\rho}_{\text{qub}}^{(1,2)})$. As reported in Fig.~\ref{fig: CORRELATIONS}, we note that, in the presence of a timing jitter error $\sigma/\tau_{\text{opt}} = 10^{-2}$ (Figs.~\ref{fig: CORRELATIONS}a and \ref{fig: CORRELATIONS}b), the sequential protocol based on a Fock state cavity (Fig.~\ref{fig: CORRELATIONS}a) is almost unaffected by qubit-qubit emergent correlations. These are indeed negligible compared to those arising when a Gaussian state is employed (Fig.~\ref{fig: CORRELATIONS}b). This could partly explain why a Gaussian state, such as the squeezed coherent one, performs worse than a Fock state. Part of the energy transferred from the Gaussian state does not flow into the qubit populations; rather, it can contribute, for instance, to its coherences, which cannot be accessed by a population measurement.

As expected, by increasing the error to $\sigma/\tau_{\text{opt}} = 10^{-1}$ (Figs.~\ref{fig: CORRELATIONS}c and \ref{fig: CORRELATIONS}d), we observe non-negligible correlations even in the Fock state protocol (Fig.~\ref{fig: CORRELATIONS}c), whose performance is indeed diminished. The correlation magnitude is related to the timing jitter error amplitude, as is the case for the Gaussian state (Fig.~\ref{fig: CORRELATIONS}d). Therefore, we note that, in general, a charging strategy based on a Fock state cavity allows the suppression of qubit-qubit correlations, resulting in better performance compared to protocols based on other cavity states. Furthermore, if a Fock state cavity is exploited, the effects of spurious correlations arising from a suboptimal choice of the interaction times can be strongly mitigated by reducing the uncertainty in timing control; on the contrary, this is not possible for a protocol based on a Gaussian state.

Given that the presence of correlations can imply the emergence of nonzero coherences in the qubit state, a legitimate question arises: What are the effects of the two-point (projective) measurements assumed by the Full Counting Statistics (FCS) on the charging protocol? It is known that a projective measurement performed in a certain basis, such as the one represented by the eigenvectors of $\hat{\sigma}_z$ (referred to as ``along the $z$-axis''), destroys the coherences (i.e., the off-diagonal elements) in that basis. To answer this question, we point out that, at the beginning of the charging process of the $j$-th qubit, its state is factorized from those of all the other qubits in the stack. Therefore, the first projective measurement performed on it along the $z$-axis will not affect its state at all, since the qubit coherences are already equal to zero. The second measurement along the $z$-axis, performed after the interaction with the cavity, actually destroys any nonzero qubit coherences: but this is completely acceptable, because we are interested in the energy stored in the qubit populations, not in its coherences. The energy that might have flowed into the qubit coherences can thus be reasonably regarded as lost or, at least, inaccessible. The use of the FCS is therefore made in a proper and accurate way, and correctly accounts for the behavior of the physical quantities of interest. \\

\subsection{Cavity dissipation}
\label{app: Cavity dissipation}
Another possible limiting factor for the proposed protocol is the energy leakage from the cavity, which can be detrimental to the coherence of its own state. To address this issue, we adopted a standard open-systems point of view \cite{landi2024current_fluctuations}. We imagined that the cavity is put in contact with a thermal bath at temperature $T$, which causes the cavity dissipation with decay rate $\gamma$. Again, to frame this effect in as plausible a scenario as possible, we based our analysis on realistic parameters. In particular, following Ref.~\cite{khanahmadi2023multimode_FOR_DECAY_RATE}, we adopted $\gamma/\omega_{\text{qub}} = 10^{-3}$, which is consistent with a decay rate of $\gamma = 1$ MHz in a superconducting device, where the characteristic frequencies $\omega_{\text{qub}}$ and $\omega_{\text{cav}}$ are typically on the order of $\sim$ GHz. On this basis, we exploited the well-known Lindblad master equation to describe the effective cavity-bath interaction:
\begin{equation}
    \label{eq: Lindblad}
\frac{d}{dt}\hat{\rho}(t) = -\frac{i}{\hbar}[\hat{H}, \hat{\rho}(t)] + \gamma(\bar{n}+1)\mathcal{D}[\hat{a}]\hat{\rho}(t) + \gamma\bar{n}\mathcal{D}[\hat{a}^\dagger]\hat{\rho}(t)
\end{equation}
with $\mathcal{D}[\hat{L}]\hat{\rho} = \hat{L}\hat{\rho}\hat{L}^\dagger - \frac{1}{2}\{\hat{L}^\dagger\hat{L}, \hat{\rho}\}$, and $\hat{H}$ being the Jaynes-Cummings or the Tavis-Cummings Hamiltonian, depending on the model we are considering. The jump operators $\hat{L}$ are represented by the cavity annihilation and creation operators, $\hat{a}$ and $\hat{a}^\dagger$. Note that $\bar{n}$ is the average number of thermal photons in the bath, defined as $\bar{n} = (e^{\beta \hbar \omega_{\text{bath}}} - 1)^{-1}$, with $\beta = 1/k_B T$. In particular, we chose a characteristic bath frequency of $\omega_{\text{bath}} = 1$ GHz and a temperature $T \simeq 7$ mK such that $\bar{n} \simeq 0.54$. As mentioned in the main text, we can formulate an equation analogous to Eq.~(\ref{eq: Lindblad}) in terms of the tilted density matrix $\hat{\rho}_\chi(t)$ and the tilted Hamiltonian $\hat{H}_\chi$ to calculate the statistics of the qubit-cavity exchanged energy (even in the presence of dissipative terms). We solved this equation for the specific decay rate $\gamma/\omega_{\text{qub}} = 10^{-3}$, but we additionally compared the effects it has on the Fock state performance by varying its magnitude between $10^{-3}$ and $10^{-7}$. Again, to emphasize that the model does not need to be based on a perfect resonance condition, we performed these simulations in an off-resonance regime defined by $\Delta \omega/\omega_{\text{qub}} = 10^{-3}$.

\begin{figure}[htp!]
    \centering
    \includegraphics[width=0.9\textwidth]{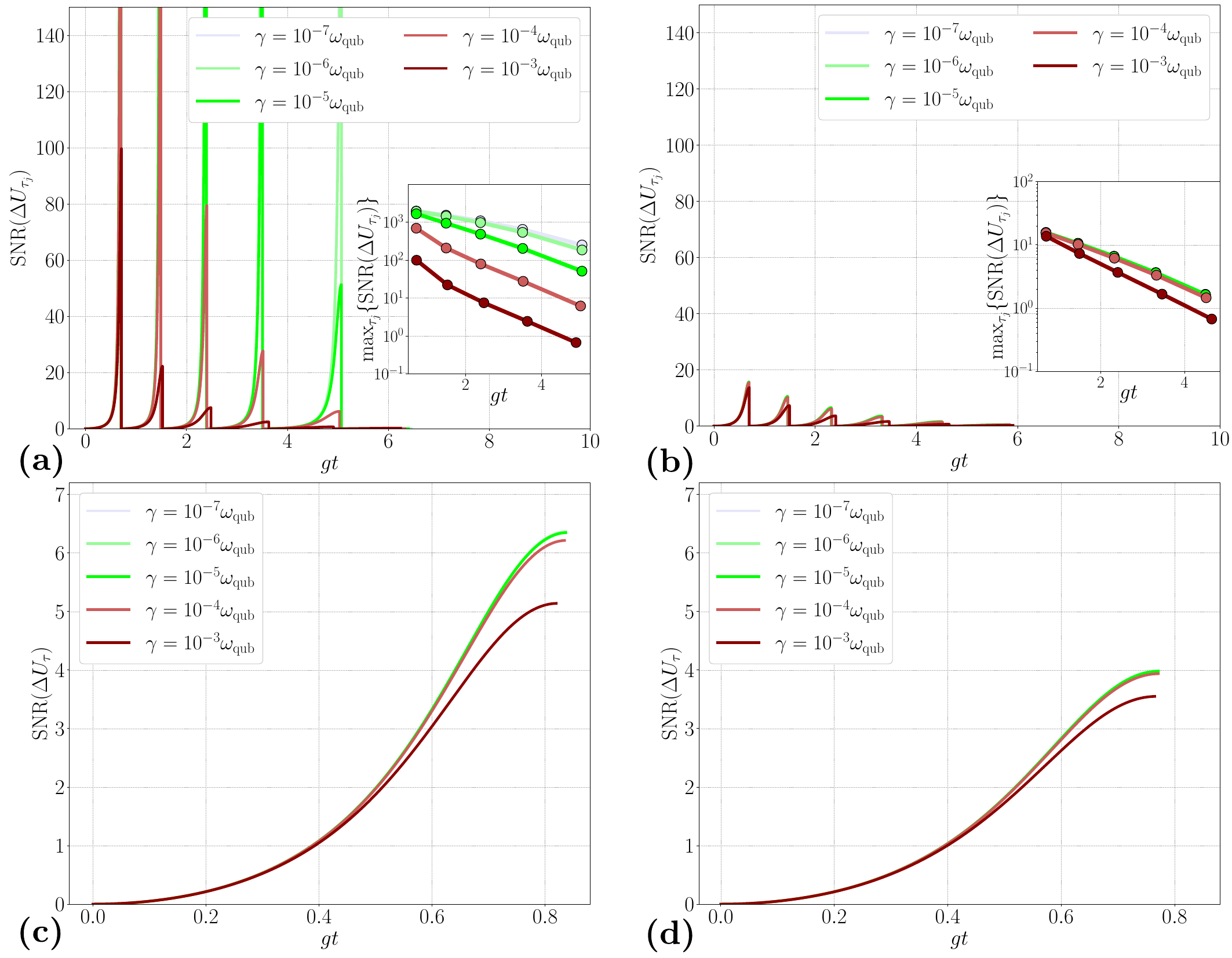}
    \caption{Accounting for cavity dissipation in a $M=5$ qubits charging process, where the cavity decay rate $\gamma/\omega_{\text{qub}}$ is varied from $10^{-7}$ to $10^{-3}$. (a) Fock state cavity employed in a sequential charging protocol. (b) Gaussian state in a sequential protocol. In (a) and (b), the inset show the maximum achieved SNR per round. (c) Parallel charging protocol with a Fock state. (d) Parallel protocol with a Gaussian state. Note that we maintained the same scale on the vertical axis for Fig.s~(a-b) and for Fig.s~(c-d), to better compare the performances of different cavity states within the same charging strategy. For these simulations, we used the same parameters of Fig.~\ref{fig: TIME JITTER 1}. We also emphasize that these results have been obtained in a slightly off-resonance regime: $\Delta\omega/\omega_{\text{qub}}=10^{-3}$.} 
    \label{fig: DECAY 1}
\end{figure}

\begin{figure}[htp!]
    \centering
    \includegraphics[width=0.95\textwidth]{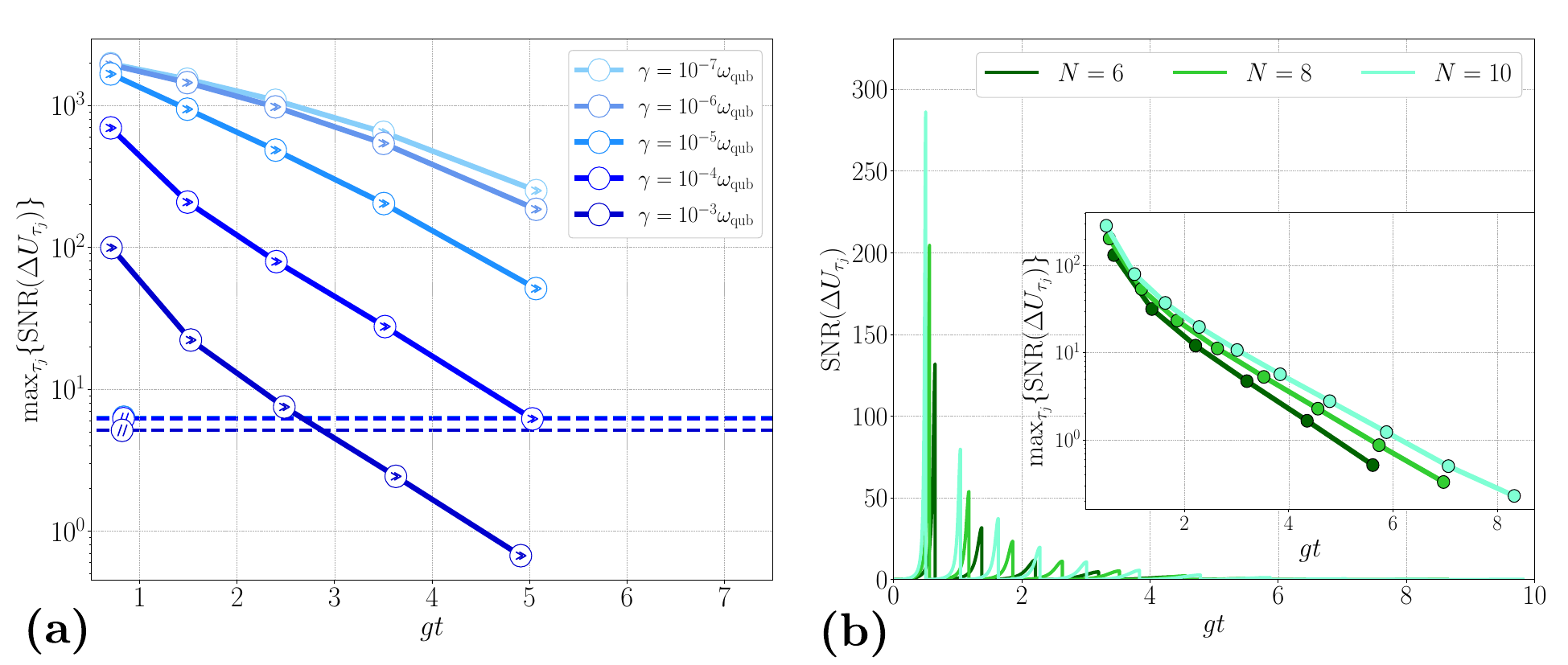}
    \caption{(a) Comparison between the performances of the $N=5$ Fock state employed in the sequential protocol and the same state used in the parallel protocol to charge $M=5$ qubits; both the protocols are subject to a decay rate $\gamma$, the values of which range between $\gamma/\omega_{\text{qub}} = 10^{-7}$ and $\gamma/\omega_{\text{qub}} = 10^{-3}$. Each point marked by ``$\gg$'' represents the maximum achieved SNR per round in the sequential protocol, while the points denoted with ``$//$'' indicate the maximum SNR obtained for each single qubit in the parallel protocol. Additional dashed lines are inserted as a guide to the eyes, to indicate the SNR that can be reached by the parallel charging strategy. The performances of the sequential protocol and of the parallel one become comparable only for $\gamma/\omega_{\text{qub}} \geq 10^{-4}$, while charging the last qubits. For $\gamma/\omega_{\text{qub}} = 10^{-3}$, the parallel protocol provides a higher SNR than the sequential protocol when charging the 4th and the 5th qubits. For $\gamma/\omega_{\text{qub}} = 10^{-4}$, the sequential protocol always performs better, except for the 5th qubit charging, for which the performances of sequential and parallel strategy are practically the same. For $\gamma/\omega_{\text{qub}} < 10^{-4}$, the sequential protocol allows to achieve the highest precision on the charge of all the qubits. (b) Scaling of the detrimental effect on the Fock state sequential protocol with the number of photons $N$ in presence of a fixed decay rate $\gamma/\omega_{\text{qub}} = 10^{-3}$. States of $N$ photons can charge up to $N-1$ qubits with more precision than states of $N-1$ photons, but the charging precision on the $N$-th qubit decreases as $N$ increases.} 
    \label{fig: DECAY 2}
\end{figure}

As depicted in Fig.~\ref{fig: DECAY 1}a, we note that the Fock state sequential protocol is expectedly limited by the presence of realistic dissipation dictated by a decay rate $\gamma/\omega_{\text{qub}} = 10^{-3}$. However, the same happens to the Gaussian state sequential protocol (Fig.~\ref{fig: DECAY 1}b), which for short times (approximately the protocol duration to charge 2-3 over 5 qubits) is still less performing than the Fock state procedure. Also, in the parallel case (Fig.~\ref{fig: DECAY 1}c for the Fock state and Fig.~\ref{fig: DECAY 1}d for the Gaussian state), we note that the SNR performances are still much lower than those of the sequential protocol, although becoming comparable when the last qubits are charged. In Fig.~\ref{fig: DECAY 2}a, we compare the performances of the Fock state employed in the sequential protocol with respect to the same state used in the parallel protocol, with a decay rate ranging between $\gamma/\omega_{\text{qub}} = 10^{-7}$ and $\gamma/\omega_{\text{qub}} = 10^{-3}$. Here, we note that the sequential strategy is more performing than the parallel one at least for the first 3 qubits over 5. Then, in correspondence with the 4th qubit, the sequential protocol performs worse than the parallel one. However, we note that if the dissipation rate is reduced to $\gamma/\omega_{\text{qub}} = 10^{-4}$ instead of $\gamma/\omega_{\text{qub}} = 10^{-3}$, then the sequential protocol performs better than the parallel one for 4 qubits over 5; for the last qubit, we find that the two protocols give almost identical results. For $\gamma/\omega_{\text{qub}} < 10^{-5}$, the sequential protocol is more advantageous than the parallel one for each of the 5 qubits. Finally, we illustrate in Fig.~\ref{fig: DECAY 2}b how the detrimental effect on the Fock state sequential protocol scales with the number of photons $N$ in the presence of a fixed decay rate $\gamma/\omega_{\text{qub}} = 10^{-3}$. For clarity, we show here the trend for $N=6,8,10$ only. We find that states of $N$ photons can charge up to $N-1$ qubits with more precision than states of $N-1$ photons. Interestingly, the performances on the last qubit charging worsen as $N$ increases. This can be explained as follows: as $N$ increases, the charging times $\tau_j$ of the first qubits are shorter, and the latter are consequently less affected by the dissipation. However, since the number of qubits to be charged is higher, the last ones will be coupled to the resonator later and therefore will suffer from a more disrupted cavity state. \\

\par\noindent
Given these results, we confirm as it was expected that the presence of energy leakage from the cavity is one of the most impacting issues on the Fock-state sequential protocol performance. Nevertheless, we also found that by reducing the dissipation, the precision improvement is such that the sequential protocol can still be more advantageous than strategies involving other kinds of states or a parallel charging mechanism. Furthermore, by increasing the number of photons $N$, the sequential protocol can prove itself increasingly more robust while charging up to $N-1$ qubits. Finally, we note that the protocol could be refined by resetting the cavity after $N-1$ charging rounds to obtain a fresh Fock state and simultaneously maintain an acceptable level of precision.

\clearpage
\section{Extensions to the Jaynes-Cummings model}
\label{app: Extension to the JC model}
Now, we address the task of investigating models that are slightly more complicated than the Jaynes-Cummings paradigm, yet focusing on the single-qubit ($M=1$) case. In particular, we first consider a time-dependent  Jaynes-Cummings interaction (Extension 1), which makes the qubit-field coupling more realistic. Then, we account for the effects of additional terms in the Hamiltonian that appear in the Rabi model (Extension 2). 

To further motivate the fact that the Fock state advantage can not be merely brought back to a perfect exchange of a precise number of photons, we thereby introduce two additional model generalizations which, analogously to what happens in the Tavis-Cummings model, can not be viewed as a sequence of swaps of energy quanta. To account for such more complex dynamics, we start from the basic Jaynes-Cummings Hamiltonian $\hat{H}_{\text{JC}}$: the model is extended by adding new specific terms to it. In particular, we investigate the anisotropic Rabi model (Extension 3), where the same counter-rotating terms of the Rabi Hamiltonian are present with different magnitudes, and a Jaynes-Cummings model with second-order nonlinear interactions (Extension 4). For both the situations, we analyze the sequential charging process of $M=5$ qubits.

\subsection{Extension 1: Time-dependent interaction}
\label{sec: Extension 1: Time-dependent interaction}
\begin{figure}[htp!]
    \centering
    \includegraphics[width=0.48\textwidth]{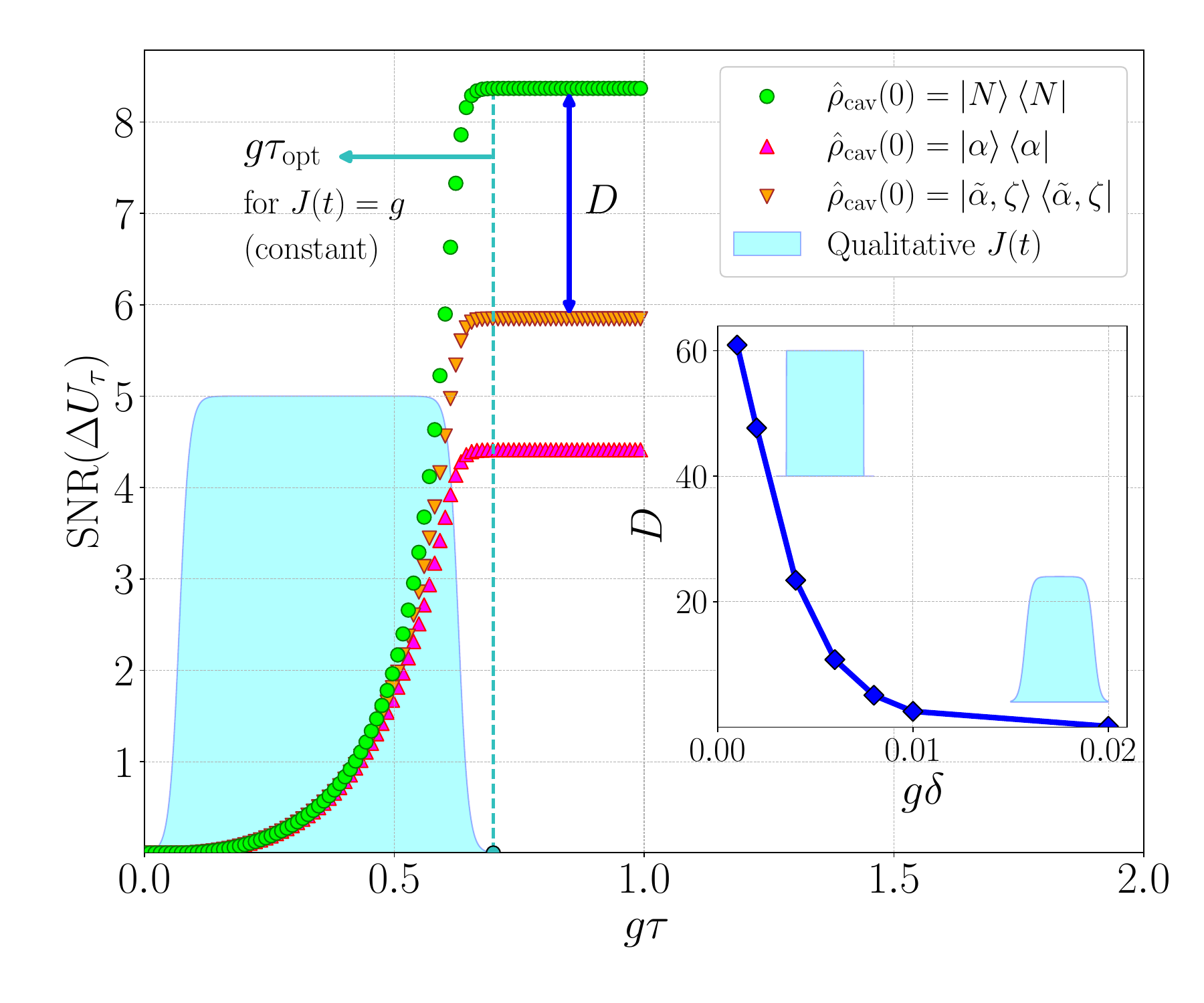}
    \caption{Example of non-uniform coupling strength $J(t)$ over time. Here, we modelled $J(t)$ as a smoothed square function (\ref{eq: Smooth J(t)}) that is almost zero at the starting time $\tau = 0$ and at the final time $\tau_{\text{opt}}$. As we can see, the advantage of the Fock state cavity over the other cavity states is still present, even if lowered if compared to the basic JC model case. The performances can be improved by approximating the coupling strength $J(t) \simeq g$ $\forall \tau\in[0, \tau_{\text{opt}}]$ as much as possible, as shown in the inset. Here, indeed, we can see that the difference $D$ between the maximum SNR for the Fock state cavity and the one for the squeezed coherent state cavity increases as $g\delta \to 0$, so for a square-like $J(t)$ (here, $g/\omega_{\text{qub}} = 10^{-2} $ is maintained fixed, while $\delta$ is varied). For that simulation, we used $\langle n \rangle = N=5$, $\alpha = \sqrt{5}$ for the coherent state, $\zeta = 0.6$ and $\tilde{\alpha} \simeq 3.905$ for the squeezed coherent state, $g/\omega_{\text{qub}} = 10^{-2}$, and $\Delta\omega = 5 \cdot 10^{-3}$. Besides, the parameters from Eq.~(\ref{eq: Smooth J(t)}) in the main plot have been chosen as $g\delta = 10^{-2}$ and $\tilde{t} = \tau_{\text{opt}}/2$. In that way, $t_1^{(\delta)} = 0.1 \tau_{\text{opt}}$ and $t_2^{(\delta)} = 0.8 \tau_{\text{opt}}$.
    }
    \label{fig: FIG2}
\end{figure}

As a first extension, we relax the hypothesis of constant coupling $g$. We thus consider a time-dependent JC model for $M=1$, where the time dependence $A_1(t)$ is replaced by a less trivial function $J(t)$ that modulates the interaction Hamiltonian:
\begin{equation}
    \label{eq: Interaction time-dependent Hamiltonian}
     \hat{H}_{\text{int}}(t)
    = \hbar J(t) (\hat{\sigma}_+\hat{a} + \hat{\sigma}_-\hat{a}^\dagger)
\end{equation}
This feature can be used to define more realistically the side effect of a finite-size cavity: we may think, for instance, of an e.m. field that is weaker near the cavity edges, but that is stronger and uniform near the center. We model such situation via a smoothed square function:
\begin{equation}
    \label{eq: Smooth J(t)}
    J(t) = \begin{cases} g[1+e^{-(t-t_1^{(\delta)})/\delta}]^{-1} \quad \text{if\quad} t < t_1^{(\delta)} + 10\delta \\ 
    g \quad\quad\quad\quad\quad\,\,\, \text{if\quad}  t \in [t_1^{(\delta)} + 10\delta, t_2^{(\delta)} - 10\delta] \\ 
     g[1+e^{+(t-t_2^{(\delta)})/\delta}]^{-1} \quad\, \text{if\quad} t > t_2^{(\delta)} - 10\delta \\
    \end{cases}
\end{equation}
where $t_1^{(\delta)}, t_2^{(\delta)}$ are such that $t_1^{(\delta)} = 20g\delta \tilde{t}$ and $t_2^{(\delta)} = 2 \tilde{t} -20g\delta \tilde{t}$, being $\tilde{t}$ the time reference with respect to which the function $J(t)$ is symmetric. In particular, we choose the parameters $\tilde{t} = \tau_{\text{opt}}/2$ and $\delta$ such that $J(t=0) \simeq 0 \simeq J(t=\tau_{\text{opt}})$, being $\tau_{\text{opt}}$ the optimal time for which a perfect charging can be obtained by using a Fock state cavity with a constant coupling strength $g$ \cite{rinaldi2025reliable}. As reported in Fig.~\ref{fig: FIG2}, a precision advantage of the Fock state cavity over the other states can still be obtained, and it can be arbitrarily enhanced as long as $J(t)$ can approximate the constant value $g$, that is, when $J(t) \simeq g$ for $\tau\in{[0, \tau_{\text{opt}}]}$, and $J(t) = 0$ otherwise. \\

\par\noindent
Since in the present situation the JC Hamiltonian is no longer time-independent, we can not use the analytical expression of the JC evolution operator $\hat{\mathcal{U}}_{\text{JC}}(t,0)$ anymore. So, as before, we resort to a numerical solution of the differential equation for the tilted density matrix $\hat{\rho}_\chi(t)$, which reads \cite{landi2022nonequilibrium}:
\begin{equation}
\label{eq: Differential equation for rho_chi}
    \frac{d}{dt}\hat{\rho}_{\chi}(t) = -\frac{i}{\hbar}\bigl[ \hat{H}_{\text{JC},\chi}(t) \hat{\rho}_{\chi}(t) - \hat{\rho}_{\chi}(t) \hat{H}_{\text{JC}, -\chi}(t) \bigr]\\
\end{equation}
where $\hat{H}_{\text{JC}, \chi}(t)$ is the tilted JC Hamiltonian $\hat{H}_{\text{JC}, \chi}(t) = \hat{H}_{\text{free}} + \hat{H}_{\text{int}, \chi}(t)$, defined as:
\begin{equation}
    \label{eq: Tilted Hamiltonian}
    \begin{split}
             \hat{H}_{\text{int}, \chi}(t) &= e^{i\frac{\chi}{2}\hat{H}_{\text{qub}}} \hat{H}_{\text{int}}(t) e^{-i\frac{\chi}{2}\hat{H}_{\text{qub}}} \\
             &= \hbar J(t)( \hat{\sigma}_+\hat{a}e^{i\frac{\chi}{2}\hbar\omega_{\text{qub}}} +  \hat{\sigma}_-\hat{a}^\dagger e^{-i\frac{\chi}{2}\hbar\omega_{\text{qub}}} )
    \end{split}
\end{equation}
and $\hat{H}_{\text{free}} = \hat{H}_{\text{qub}} + \hat{H}_{\text{cav}}$.

\subsection{Extension 2: Isotropic Rabi model}
\label{sec: Extension 2: Rabi model}
The second model extension goes beyond the rotating-wave approximation (RWA): in that case, the JC paradigm changes its name into the Rabi model \cite{braak2019symmetries}, whose Hamiltonian is:
\begin{equation}
    \label{eq: Rabi model}
    \hat{H}_{\text{Rabi}} = \hbar\omega_{\text{qub}} \frac{\hat{\sigma}_z}{2} + \hbar\omega_{\text{cav}} \hat{a}^\dagger  \hat{a} + \hbar g \hat{\sigma}_x( \hat{a} +  \hat{a}^\dagger)
\end{equation}
Note that, for simplicity, we are restricting ourselves to the case of a constant coupling strength, $J(t) = g$. In the adopted qubit basis notation, $\hat{\sigma}_x( \hat{a} +  \hat{a}^\dagger)$ can be rewritten as $\hat{\sigma}_+ \hat{a} +  \hat{\sigma}_+\hat{a}^\dagger \hat{\sigma}_- \hat{a} +  \hat{\sigma}_-\hat{a}^\dagger$. The counter-rotating terms (CRT) are the ones represented by $\hat{\sigma}_+\hat{a}^\dagger$ and $\hat{\sigma}_- \hat{a}$. As mentioned before, such terms are usually neglected in the weak-coupling regime via the RWA. These are not energy-preserving terms; however, if the model is written in interaction picture, they turn out to be weighted by complex phases as $e^{\pm i(\omega_{\text{qub}} + \omega_{\text{cav}})t}$, which mediate to zero as time flows (thus making the CRT vanishing), unless that $g$ is comparable to $\omega_{\text{qub}}$ and $ \omega_{\text{cav}}$. In that case, these terms are no longer negligible and they should be taken into account.  \\
\begin{figure}[htp!]
    \centering
    \includegraphics[width=0.9\textwidth]{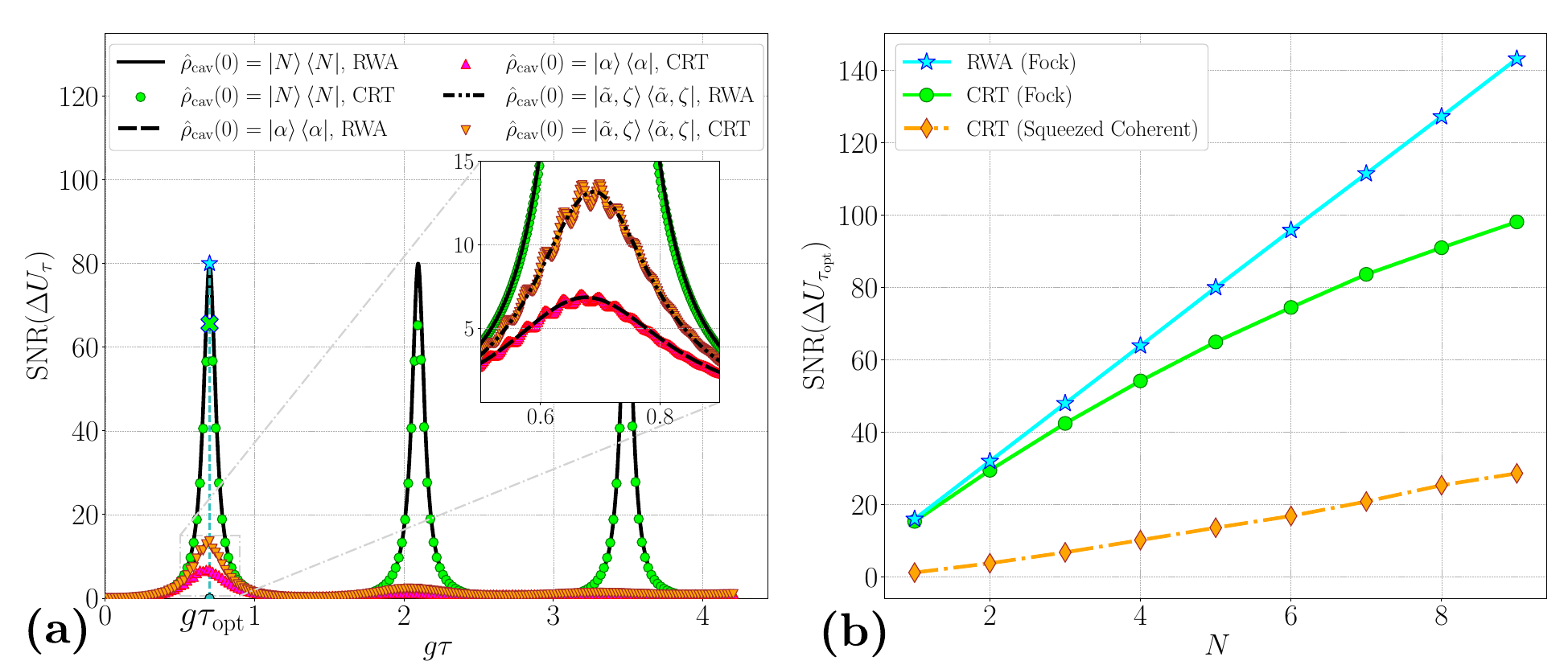}
    \caption{Inclusion of the counter-rotating terms (CRT) $\hat{\sigma}_-\hat{a}$ and $\hat{\sigma}_+\hat{a}^\dagger$ in the JC Hamiltonian. The quantum battery is now described by the Rabi model. (a) Dynamics of the single-qubit evolution. The black lines represent the basic model, i.e., JC in rotating-wave approximation (RWA). The discrete points show the results for the Rabi model. Although when including the CRT the maximum of the Fock state SNR is lowered (green cross) with respect to the maximum in the RWA case (light blue star), a consistent advantage can still be observed. Besides, small oscillations in the SNR arise for the coherent and squeezed coherent states. (b) Analysis of the error accumulation due to the effect of the CRT on the single-qubit charging process, varying the initial number of photons $N = \langle n \rangle$ in the cavity. Here, each point corresponds to the maximum SNR obtained for the qubit charging. As it can be seen, the discrepancy between the RWA model (blue stars) and the CRT  model (green bullets) is effectively increasing as $N$ increases; however, the performances of the Fock state in presence of the CRT are still far above the ones of the squeezed coherent state (orange diamonds). For that simulation, we used $\langle n \rangle = N=5$, $\alpha = \sqrt{5}$ for the coherent state, $\zeta = 0.6$ and $\tilde{\alpha} \simeq 3.905$ for the squeezed coherent state, $g/\omega_{\text{qub}} = 10^{-2}$, and $\Delta\omega/\omega_{\text{qub}} = 5 \cdot 10^{-3}$. 
    }
    \label{fig: FIG3}
\end{figure}

The effects of the CRT are reported in Fig.~\ref{fig: FIG3}, which shows the numerical results obtained by finding the tilted evolution operator numerically. As expected, they can be limiting for the Fock state cavity advantage, since the maximum SNR can be no longer arbitrarily enhanced. Furthermore, we note that the error made by performing the RWA accumulates as the number of photons $N$ in the Fock state becomes large. However, the SNR obtained through the use of a Fock state still largely outperforms the one computed for other cavity states. Nevertheless, we conclude that, in order to have a precision-focused quantum battery, the optimal regime in which to design the device is the JC one, i.e., the one for which qubit and cavity are weakly coupled (i.e., $g/\omega_{\text{qub}} \ll 1$), and thus the CRT can be safely neglected.

\subsection{Extension 3: Anisotropic Rabi model}
\label{app: Extension 3: Anisotropic Rabi model}
\begin{figure}[htp!]
    \centering
    \includegraphics[width=0.9\textwidth]{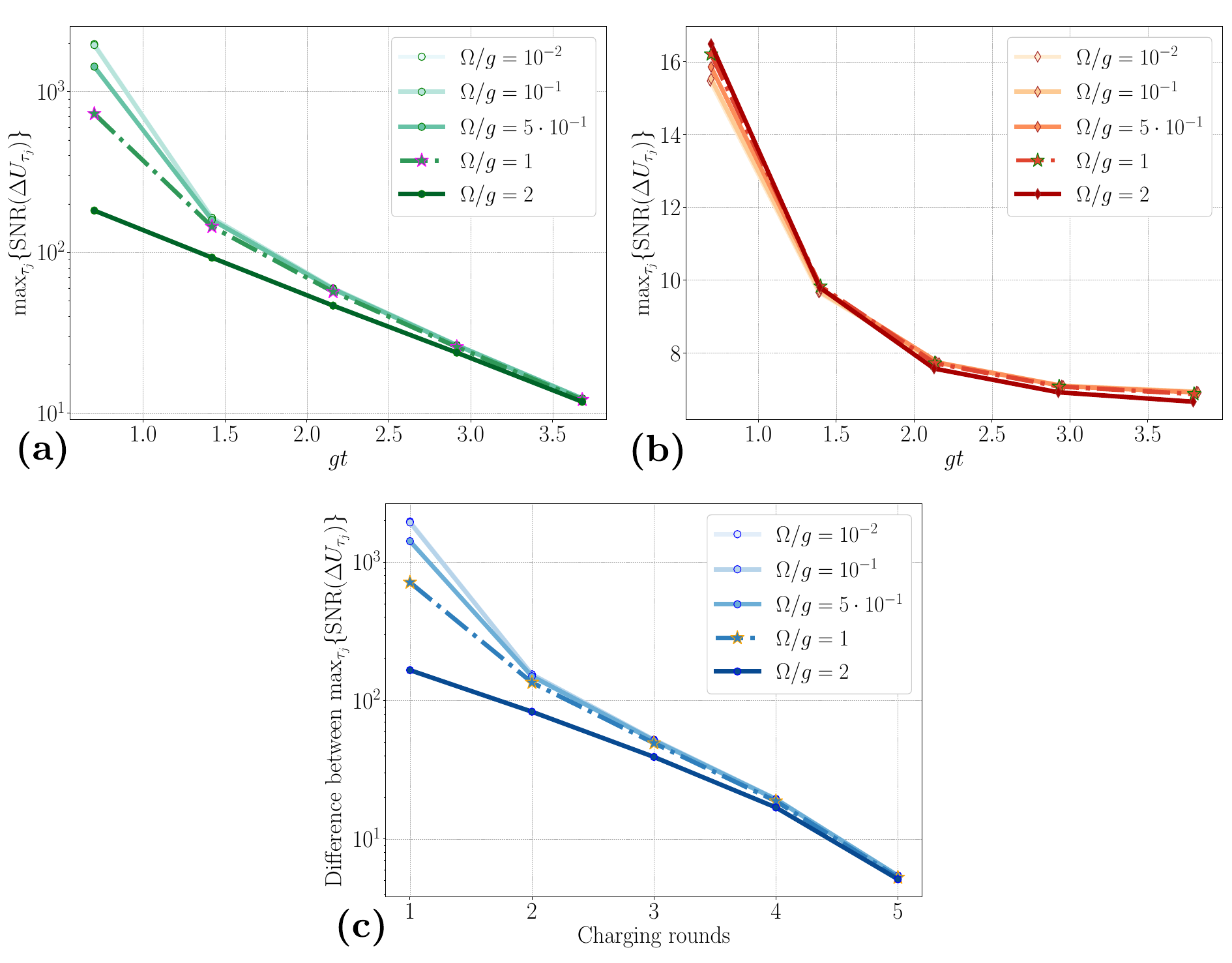}
    \caption{Anisotropic Rabi model, where counter-rotating terms with increasing strength $\Omega$ are present. The plots show the maximum achieved SNR for each charged qubit in the charging protocol of $M=5$ qubits, by employing (a) a Fock cavity state and (b) a squeezed coherent cavity state. (c) Difference between the SNR achieved by the Fock state protocol and the one achieved by the Gaussian state protocol. We note that, although the difference is diminished at long times, it always remains positive. For clarity, the points associated with the Rabi model (i.e., $\Omega/g = 1$) are marked by a star and a dash-dotted line. The simulation parameters are the same as in Fig.~\ref{fig: FIG3}, except for the detuning, which now is $\Delta\omega/\omega_{\text{qub}} =  10^{-3}$.} 
    \label{fig: Intermediate Rabi}
\end{figure}

At this point, we can further extend our analysis through the introduction of unbalanced counter-rotating terms, which constitute the peculiarity of the so-called anisotropic Rabi model \cite{xie2014anisotropicRabi}. Such new terms, modulated by the constant factor $\Omega$, are added to the basic JC Hamiltonian as follows:
\begin{equation}
\hat{H}_{\text{CRT}} = \hat{H}_{\text{JC}} + \hbar\Omega (\hat{\sigma}_-\hat{a} + \hat{\sigma}_+\hat{a}^{\dagger})
\end{equation}
These additional elements, represented by `non-energy preserving' operators $\hat{\sigma}_-\hat{a} + \hat{\sigma}_+\hat{a}^{\dagger}$, are the same that appear in the (isotropic) Rabi Hamiltonian, which here is obtained when $\Omega/g = 1$. Here, however, we explore the intermediate regimes that originate when $\Omega$ has magnitude comprised between $\Omega/g = 10^{-2}$ and $\Omega/g = 2$, so even superior to the one provided by the Rabi Hamiltonian. We simulated the sequential charging of $M=5$ qubits with the influence of the counter-rotating terms. Here and in what follows, we also consider an off-resonance condition $\Delta\omega/{\text{qub}} = 10^{-3}$ to avoid idealized SNR divergences. Our findings are summarized in Fig.~\ref{fig: Intermediate Rabi}, from which we can note that the Fock state SNR (Fig.~\ref{fig: Intermediate Rabi}a), if compared to the one provided by a Gaussian state (Fig.~\ref{fig: Intermediate Rabi}b), remains higher even for non-negligible magnitudes of $\Omega$. If, on one hand, the Fock state performances are diminished over long times, on the other hand they remain superior to the ones of the Gaussian state: their difference (Fig.~\ref{fig: Intermediate Rabi}c), indeed, is always positive.  \\

\subsection{Extension 4: Nonlinear interactions}
\label{app: Extension 4: Nonlinear interactions}
\begin{figure}[htp!]
    \centering
    \includegraphics[width=0.9\textwidth]{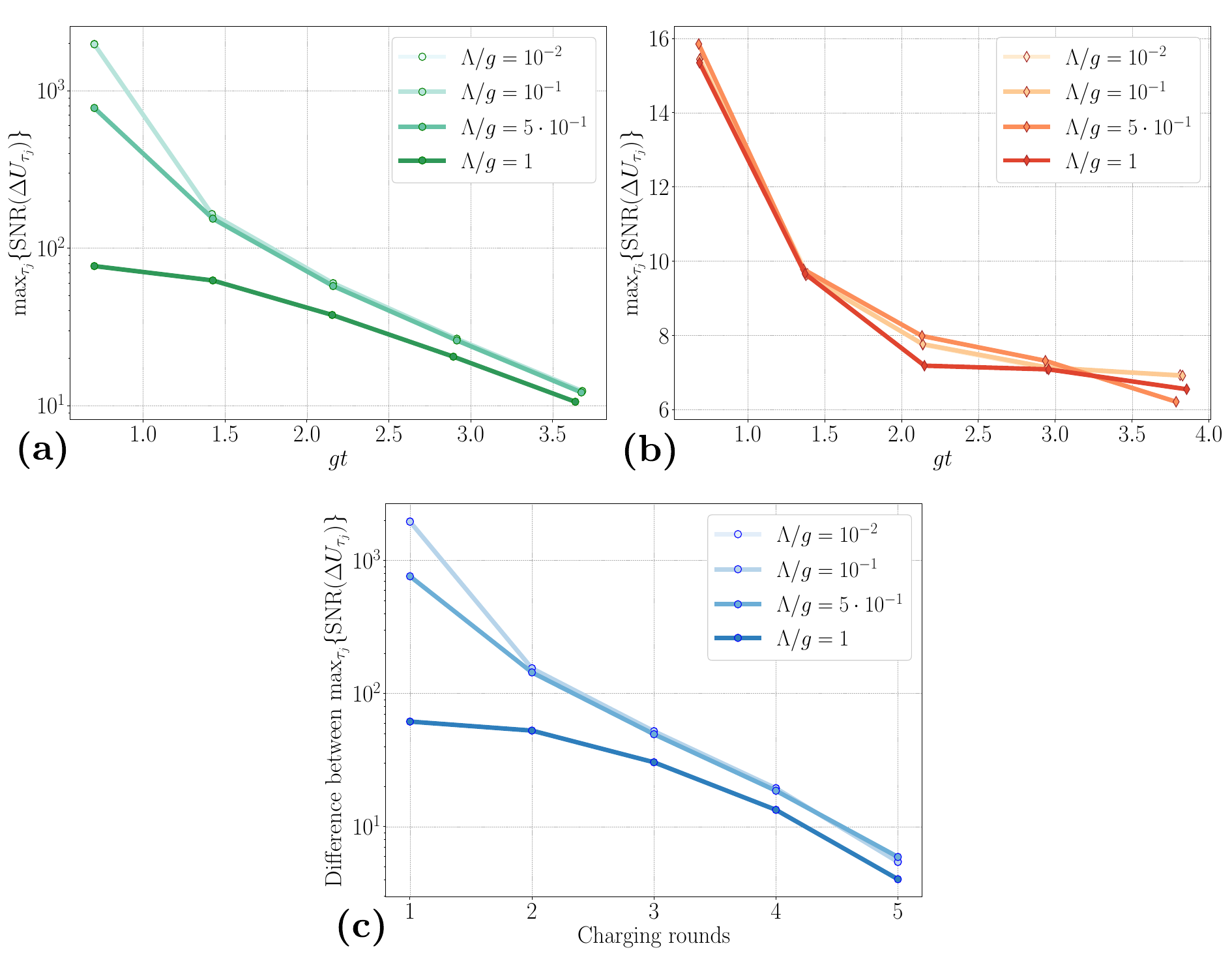}
    \caption{Jaynes-Cummings model in presence of nonlinear interaction terms, such as $\hat{\sigma}_+\hat{a}^2 + \hat{\sigma}_-\hat{a}^{\dagger2}$, with increasing magnitude $\Lambda$. Each plot represents the maximum achieved SNR for each charged qubit in the charging protocol of $M=5$ qubits, where (a) a Fock cavity state or (b) a squeezed coherent cavity state is employed. The difference between the SNR achieved by the Fock state protocol and the one achieved by the Gaussian state protocol is shown in (c). Again, such difference is still positive even for long times. The simulation parameters are the same as in Fig.~\ref{fig: Intermediate Rabi}.} 
    \label{fig: Nonlinearity}
\end{figure}

Another model generalization is represented by the inclusion of higher-order nonlinear interaction terms. Drawing inspiration from \cite{andolina2025genuine_anharmonic}, we accounted for the presence of nonlinear interactions mediated by quadratic field operators, and modulated by a coupling factor $\Lambda$. The resulting total Hamiltonian is:
\begin{equation}
\hat{H}_{\text{NL}} = \hat{H}_{\text{JC}} + \hbar\Lambda \bigl(\hat{\sigma}_+\hat{a}^2 + \hat{\sigma}_-\hat{a}^{\dagger2}\bigr)
\end{equation}
where the coupling strength $\Lambda$ has been allowed to vary between $\Lambda/g = 10^{-2}$ and $\Lambda/g = 1$. We report our results in Fig.~\ref{fig: Nonlinearity}, showing the sequential charging of $M=5$ qubits in presence of nonlinear interactions. Similarly to what was illustrated above, we observe that the Fock state protocol (Fig.~\ref{fig: Nonlinearity}a), although negatively influenced by the presence of this kind of interaction over long times, is still better performing than its Gaussian counterpart (Fig.~\ref{fig: Nonlinearity}b). The difference between the two (Fig.~\ref{fig: Nonlinearity}c) increases as the magnitude $\Lambda$ of the nonlinear interaction decreases; however, for smaller values of $\Lambda/g$, such difference saturates, suggesting that the presence of weak nonlinear terms has no relevant impact on the protocol performances. \\

Such an analysis allows us to conclude that the effects of nonlinear interaction terms or of non-energy preserving operators in the Hamiltonian might be detrimental for the performances of the sequential charging protocol: therefore, the basic Jaynes-Cummings constitutes the optimal choice to achieve the best results in terms of charging precision. Nevertheless, our results also prove that the Fock state advantage is still present even if the underlying dynamics is not strictly the one generated by the basic JC Hamiltonian. This fact further proves that, just as for the Tavis-Cummings description, these models can not be reduced to a sequence of excitation swaps. The persistence of the Fock state advantage even during the associated dynamics, however, suggests that the precision enhancement has a more profound root: it can not be explained simply by bringing it back to the number-definition of a $N$-photon Fock state. \\

\section{Phase randomized state}
\label{app: Phase randomized state}
To account for more realistic Gaussian states, we considered a phase-randomized coherent squeezed state \cite{pradana2019quantum} in place of the usual squeezed coherent state to perform the simulations related to the noise analysis. Such a state can be written as:
\begin{equation}
    \label{eq: Coherent squeezed state, from pradana2019quantum, RANDOMIZED, appendix version}
    \begin{split}
    \hat{\rho} &= \frac{1}{2\pi}\int_0^{2\pi}d\theta \sum_{n=0}^{\infty}\sum_{n'=0}^{\infty} p_n(\zeta,\tilde{\alpha})  p_{n'}(\zeta,\tilde{\alpha}) \ket{n}\bra{n'}  = \sum_{n=0}^{\infty} \mathscr{D}(n) \ket{n}\bra{n}
    \end{split}
\end{equation}
where the probability distribution over $n$ is given by:
\begin{equation}
    \label{eq: Coherent squeezed with randomized phases distribution}
    \begin{split}
     \mathscr{D}(n) = \frac{1}{n!\cosh{r}}\bigl[\frac{1}{2} \tanh{r}\bigr]^{n}e^{-\bigl( |\alpha|^2 (1 + \tanh{r}) \bigr)}  \biggl\{H_n\biggl(|\alpha|\frac{ 1+\tanh{r}}{\sqrt{2\tanh{r}}}\biggr)\biggr\}^2
    \end{split}
\end{equation}
being $H_n(x)$ the $n$-th Hermite polynomial of $x$, and $r,\phi$ such that $\zeta = re^{i\phi}$. The phase-randomized state (\ref{eq: Coherent squeezed state, from pradana2019quantum, RANDOMIZED, appendix version}) can be obtained from the explicit expression of the coherent squeezed state $\ket{\zeta,\alpha}\bra{\zeta,\alpha}$ (see \cite{pradana2019quantum} for details) such that:
\begin{equation}
    \label{eq: Coherent squeezed state, from pradana2019quantum}
    \begin{split}
        \ket{\zeta,\alpha} &= \sum_{n=0}^{\infty} \frac{1}{\sqrt{n!\cosh{r}}}\bigl[\frac{1}{2}e^{i\phi} \tanh{r}\bigr]^{\frac{n}{2}}  e^{-\frac{1}{2}\bigl( |\alpha|^2 + (\alpha^\ast)^2 e^{i\phi} \tanh{r} \bigr)}  
         H_n\biggl(\frac{\alpha + \alpha^\ast e^{i\phi}\tanh{r}}{\sqrt{2e^{i\phi}\tanh{r}}}\biggr) \ket{n} \equiv \sum_{n=0}^{\infty} p_n(\zeta,\alpha) \ket{n}
    \end{split}
\end{equation}
by setting $\phi=2\theta$, then integrating over $\theta\in[0, 2\pi]$ and by dividing by $2\pi$. \\

We note that, when $\alpha$ and $\zeta$ are real numbers (and therefore $\phi = 2m\pi$, with $m$ integer), then $\mathscr{D}(n) = p_n^2$: therefore, a Gaussian randomized-phase (diagonal) state and a Gaussian state with nonzero off-diagonal elements actually share the same diagonal elements. Because of that, this choice does not affect the performances of the Gaussian state if we consider diagonal qubits states of the form $\hat{\rho}_{\text{qub}}(0) = q \ket{e}\bra{e} + (1-q) \ket{g}\bra{g}$ at the beginning of the charging process. In that case, indeed, the qubit excited level population $p_e(\tau)$ depends only on the diagonal elements of the initial cavity state, $\hat{\rho}_{\text{cav}}(0)$, which are the same in both cases. This can be easily demonstrated by looking at the JC evolution operator \cite{stenholm1973quantum, smirne2010nakajima, bocanegra2024invariant, larson2021jaynes, rinaldi2025reliable}, which we symbolically write here as $\hat{\mathcal{U}}_{\text{JC}}(\tau) = $ \begin{scriptsize}${\begin{pmatrix}
    \hat{u}_{00} & \hat{u}_{01}\\
    \hat{u}_{10} & \hat{u}_{11}
\end{pmatrix}}$\end{scriptsize}, where $\hat{u}_{ij}$ are $\tau$-dependent cavity operators. Since the joint system evolved state is $\hat{\rho}_{\text{JC}}(\tau) = \hat{\mathcal{U}}_{\text{JC}}(\tau) \hat{\rho}_{\text{JC}}(0) \hat{\mathcal{U}}_{\text{JC}}^\dagger(\tau) $, we can write the excitation probability $p_e(\tau)$ as:
\begin{equation}
    \label{eq: Qubit excitation probability, generic}
    \begin{split}
        p_e(\tau) &= \text{Tr}\{ \ket{e}\bra{e}\hat{\rho}_{\text{JC}}(\tau)\} = q  \text{Tr}\{ \hat{u}_{00}\hat{\rho}_{\text{cav}}(0)\hat{u}^\dagger_{00} \} + (1-q)  \text{Tr}\{ \hat{u}_{01}\hat{\rho}_{\text{cav}}(0)\hat{u}^\dagger_{01} \}
    \end{split}
\end{equation}
We now may consider a non-diagonal cavity density matrix $\hat{\rho}_{\text{cav}}(0) = \sum_{k,k'}\rho_{k,k'}\ket{k}\bra{k'}$. However, if we insert here the formulae for $\hat{u}_{00}$ and $\hat{u}_{01}$ (we refer to \cite{stenholm1973quantum, smirne2010nakajima, bocanegra2024invariant, larson2021jaynes, rinaldi2025reliable} for their explicit expressions) into (\ref{eq: Qubit excitation probability, generic}), we find:
\begin{equation}
    \label{eq: Qubit excitation probability, explicit}
    \begin{split}
        p_e(\tau) &=  \sum_n q  \cos^2(g\tau\sqrt{n+1})\rho_{n,n} + (1-q)  \sin^2(g\tau\sqrt{n+1})\rho_{n+1,n+1} 
    \end{split}
\end{equation}
Thus, the only relevant matrix elements in $\hat{\rho}_{\text{cav}}(0)$ are the diagonal ones, because $p_e(\tau)$ depends only on them. Furthermore, since $p_e(\tau)$ is intimately connected with the qubit energy, we conclude that also $\langle \Delta U_\tau \rangle$ and $\text{SNR}(\Delta U_\tau)$ do not change if we consider phase randomized cavity states in place of non-diagonal ones (provided that $\alpha,\zeta\in\mathds{R}$, i.e., they are real numbers). \\
As a remark, we emphasize that this does not hold for the qubit coherences, which can depend on the off-diagonal elements of the cavity state. However, since we are interested in the energy stored in the qubit populations only, we can safely neglect the impact of the different cavity state on the qubit off-diagonal elements. \\

\section{Comparison with other quantum non-Gaussian states}
\label{app: Comparison with other quantum non-Gaussian states}
\begin{figure}[htp!]
    \centering
    \includegraphics[width=0.9\textwidth]{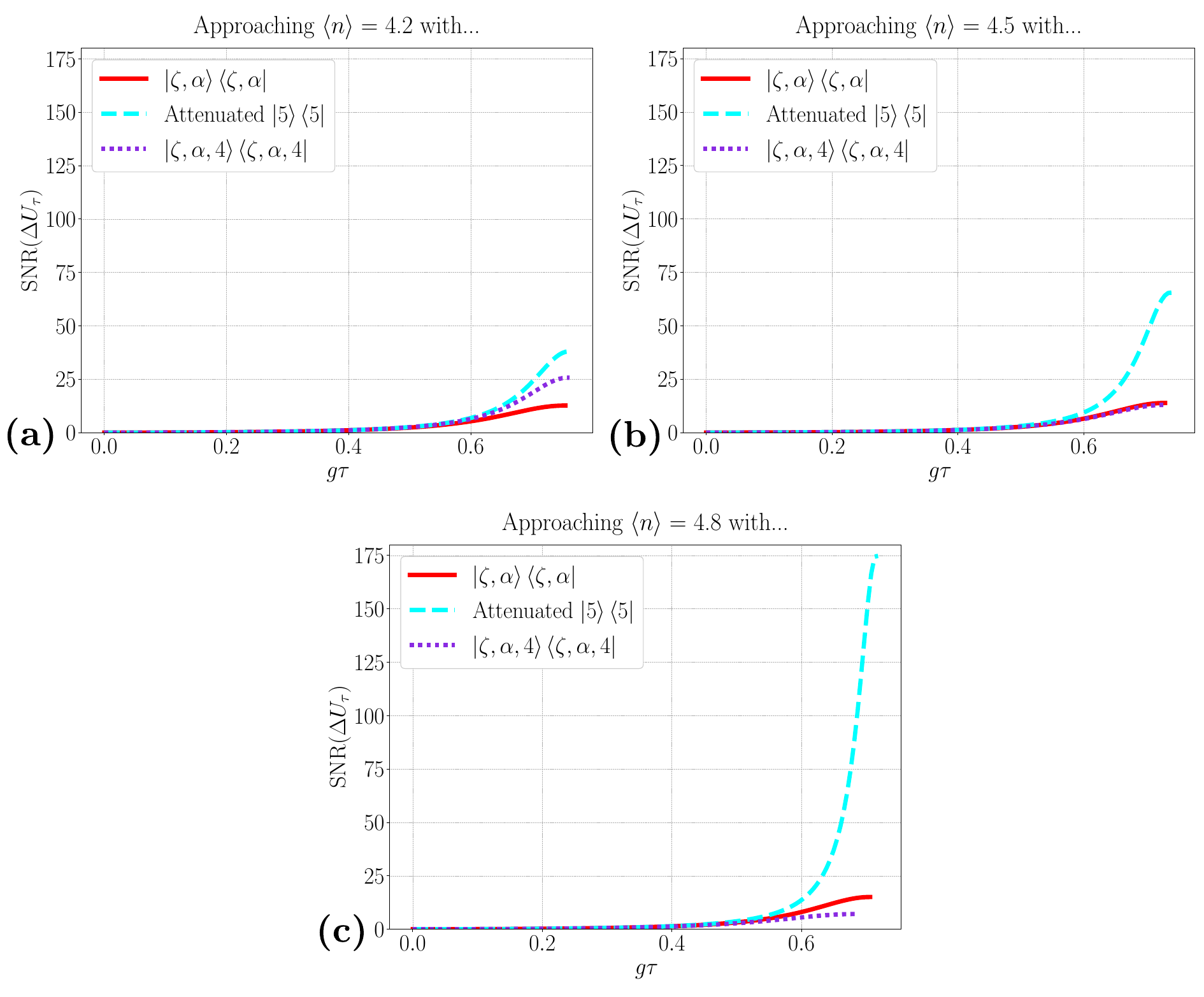}
    \caption{Performances of the single-qubit charging protocol performed with three different cavity states: a mixture of Fock states $\sum_{n=0}^{N=5} p_n \ket{n}\bra{n}$ (i.e., an attenuated Fock state $\ket{5}$), a Gaussian (coherent squeezed) state $\ket{\zeta,\alpha}\bra{\zeta,\alpha}$, and a squeezed displaced Fock state $\ket{\zeta,\alpha,4}\bra{\zeta,\alpha,4}$. The three states do not have a precise integer number of photons; rather, they have the same number of photons on average, $\langle n \rangle$. Here, we compute the SNR associated with the energy injected in the qubit over the optimal time window $g\tau$, i.e., the time at which the fidelity $F$ between the qubit state at time $t$ and the target state $\ket{e}\bra{e}$ is maximized. We consider (a) $\langle n \rangle = 4.2$, (b) $\langle n \rangle = 4.5$, and (c) $\langle n \rangle = 4.8$. }  
    \label{fig: DIFFERENT STATES}
\end{figure}

In order to interpret our results, one could wonder whether the variance suppression due to the employment of a Fock state cavity could be entirely attributed to the fact that the Fock state has, by definition, a well-defined number of photons. As a final discussion, we therefore analyzed the performance of a single-qubit charging process using three different cavity states: an attenuated Fock state (i.e., a mixture of Fock states $\sum_{n=0}^N p_n \ket{n}\bra{n}$), a Gaussian (squeezed coherent) state, and a squeezed displaced $(N-1)$-Fock state $\ket{\zeta,\alpha,N-1}\bra{\zeta,\alpha,N-1}$, where $\ket{\zeta,\alpha,N} = \hat{S}(\zeta)\hat{D}(\alpha)\ket{N}$. As evident, none of these states has a well-defined number of photons; however, we impose that they must have the same average number of photons, $\langle n \rangle$. Moreover, in our simulations, we consider a mean photon number between $\langle n \rangle = 4$ and $\langle n \rangle = 5$; specifically, we focus on the values $\langle n \rangle \in \{4.2, 4.5, 4.8\}$. Under these conditions, no photon-number advantage can be assumed \textit{a priori} (note that $\langle n \rangle$ is intentionally chosen as a non-integer). We compute the SNR associated with the energy injected into the qubit over the optimal time window $g\tau$, which is the time at which the fidelity $F$ between the qubit state at time $t$ and the target state $\ket{e}\bra{e}$ is maximized, for each cavity state. As shown in Fig.~\ref{fig: DIFFERENT STATES} in the present document, the best performance in terms of SNR is always achieved by the attenuated Fock state. The addition of squeezing and displacement to a Fock state with $N=4$, which can be viewed as a kind of `addition of Gaussianity', can increase or decrease the state's performance depending on the mean number of photons. However, in all cases, the mixture $\sum_{n=0}^N p_n \ket{n}\bra{n}$ of quantum non-Gaussian states allows for achieving maximum precision performance without having any precise initial photon number.

Experimentally, we could exploit the quantum non-Gaussian hierarchy introduced and validated in \cite{podhora2022quantum} to directly probe the cavity's initial and final states. This approach would yield deeper insights into the phenomenon and allow a more nuanced characterization of the interplay between quantumness and non-Gaussianity underlying the charging advantage.

\end{document}